\documentclass[sn-basic,Numbered,iicol]{sn-jnl}  

\usepackage{graphicx}
\usepackage{textcomp}
\usepackage{color,xcolor}
\usepackage{caption}

\usepackage{subfigure}
\usepackage{booktabs}
\usepackage{hyperref}
\usepackage[switch]{lineno}
\usepackage{soul}
\usepackage{utfsym}
\usepackage{indentfirst}
\usepackage{times}
\usepackage{latexsym}

\usepackage{graphicx}%
\usepackage{multirow}%
\usepackage{amsmath,amssymb,amsfonts}%
\usepackage{amsthm}%
\usepackage{mathrsfs}%
\usepackage[title]{appendix}%
\usepackage{xcolor}%
\usepackage{textcomp}%
\usepackage{manyfoot}%
\usepackage{booktabs}%
\usepackage{algorithm}%
\usepackage{algorithmicx}%
\usepackage{algpseudocode}%
\usepackage{listings}%

\newtheorem{theorem}{Theorem}
\newtheorem{definition}{Definition}%

\begin{document}

\title[Article Title]{GHQ: Grouped Hybrid Q-Learning for Cooperative Heterogeneous Multi-agent Reinforcement Learning}


\author[1,2]{\fnm{Xiaoyang} \sur{Yu}}\email{xiaoyang.yu@bjtu.edu.cn}
\author[1,2]{\fnm{Youfang} \sur{Lin}}
\author[1,2]{\fnm{Xiangsen} \sur{Wang}}
\author[1,2]{\fnm{Sheng} \sur{Han}}
\author*[1,2]{\fnm{Kai} \sur{Lv}}\email{lvkai@bjtu.edu.cn}

\affil[1]{\orgdiv{School of Computer and Information Technology}, \orgname{Beijing Jiaotong University}, \orgaddress{\city{Beijing}, \country{China}}}

\affil[2]{\orgdiv{Beijing Key Laboratory of Traffic Data Analysis and Mining}, \orgname{Beijing Jiaotong University}, \orgaddress{\city{Beijing}, \country{China}}}


\abstract{Previous deep multi-agent reinforcement learning (MARL) algorithms have achieved impressive results, typically in symmetric and homogeneous scenarios. However, asymmetric heterogeneous scenarios are prevalent and usually harder to solve. In this paper, the main discussion is about the cooperative heterogeneous MARL problem in asymmetric heterogeneous maps of the Starcraft Multi-Agent Challenges (SMAC) environment. Recent mainstream approaches use policy-based actor-critic algorithms to solve the heterogeneous MARL problem with various individual agent policies. However, these approaches lack formal definition and further analysis of the heterogeneity problem. Therefore, a formal definition of the Local Transition Heterogeneity (LTH) problem is first given. Then, the LTH problem in SMAC environment can be studied. In order to comprehensively reveal and study the LTH problem, some new asymmetric heterogeneous maps in SMAC are designed. It has been observed that baseline algorithms fail to perform well in the new maps. Then, the authors propose the Grouped Individual-Global-Max (GIGM) consistency and a novel MARL algorithm, Grouped Hybrid Q-Learning (GHQ). GHQ separates agents into several groups and keeps individual parameters for each group. To enhance cooperation between groups, GHQ maximizes the mutual information between trajectories of different groups. A novel hybrid structure for value factorization in GHQ is also proposed. Finally, experiments on the original and the new maps show the fabulous performance of GHQ compared to other state-of-the-art algorithms.}




\maketitle

\section{Introduction}
The multi-agent system (MAS) and multi-agent reinforcement learning (MARL) have drawn lots of attention \cite{APPI-oroojlooy2022review} and have been applied to solve some optimization problems in the physical world, such as resource allocation problem \cite{powersys+sujil2018multi,cloud+gao2020hierarchical,power+li2022dynamic}, cooperative navigation problem \cite{APPI-chen2020gama,APPI-sun2022learning,APPI-ye2022improving}, air traffic flow management \cite{APPI-kravaris2023explaining} and massive traffic light control problem \cite{APPI-qiao2023traffic,traffic+yang2021semi,traffic+liu2022distributed}. The novel research field successfully combines machine learning (ML) \cite{REF1-9979725,REF2-malakar2020ga}, deep learning (DL) \cite{REF1-Shen_2023,REF1-tao2023unsupervised}, and swarm intelligence \cite{REF2-bacanin2021performance} approaches and proves the ability to obtain outstanding results in different areas. Previous deep MARL algorithms have achieved impressive results in cooperative MARL environment \cite{hernandez2019survey,survey+gronauer2022multi}, \textit{e.g.} the Starcraft Multi-Agent Challenges (SMAC) environment \cite{smac+samvelyan2019starcraft}.\par
The SMAC environment is a multi-agent micromanagement scenario in which two adversarial MAS battle against each other. The goal is to train a MARL algorithm controlling ally agents to eliminate enemy agents controlled by the internal script of SMAC environment. The algorithm needs to learn tactics and skills for choosing the best actions and utilizing the different properties of agents. Due to the discrete property of the environment, value-based algorithms have achieved better results than policy-based algorithms \cite{qmix+rashid2018,riit+hu2021rethinking,mappo+yu2021surprising}.\par
Asymmetric heterogeneous problems are very common in real-world scenarios \cite{clauset2009power,wang2024what,wang2023skill,lv2023spatially,lv2020pose}, such as wireless network accessibility problem \cite{wifi+yu2021multi} and multi-agent robotic systems \cite{robotic+ivic2020motion,robotic+yoon2019learning}. However, the original maps of SMAC environment mainly consist of symmetric maps or homogeneous maps (see Table \ref{table-SMAC-maps}). A \textit{symmetric} map means that allies and enemies consist of the same types of units, and the numbers of both sides are also equal. A \textit{homogeneous} map means that allies consist of one specific type of unit, no matter what composition of the enemies. Furthermore, in SMAC, even though allies and enemies are equal at the starting state of symmetric heterogeneous problems, they would become asymmetric as the game runs, because the two sides are attacking and killing each other. In \cite{zhong2023heterogeneous}, the authors propose a situation (Proposition 5) where the policy of an algorithm may be trapped in a sub-optimal state due to the complexity of heterogeneity. Therefore, it is necessary to comprehensively and carefully study the heterogeneous MARL problem.\par
Previous algorithms have acquired good performance in most symmetric homogeneous maps, symmetric heterogeneous maps, and asymmetric homogeneous maps from the SMAC original map set. However, experiments show that even state-of-the-art algorithms cannot achieve a high winning-rate (\textit{WR}) in asymmetric heterogeneous maps, indicating that the combination of asymmetry and heterogeneity brings more complexity. Therefore, in order to fully study the heterogeneity problem, it is beneficial to enrich the SMAC environment with more asymmetric heterogeneous maps. Recent mainstream approaches use policy-based actor-critic algorithms to solve the heterogeneous MARL problem with various individual agent policies \cite{happo+kuba2021trust,mapg+bono2018cooperative}. Some other papers discussing heterogeneity are mainly about multi-agent robotic systems, such as \cite{robotic+ivic2020motion,robotic+yoon2019learning}, which are slightly different from MARL research. \par
For example, in the multi-agent area search problem proposed in \cite{robotic+ivic2020motion}, the multi-agent robotic methods usually manage to model the problem in detail with proper mathematical structures, and then propose the solution. However, the MARL approaches usually model the problem as a POMDP (see section \ref{Preliminaries} for details) and design a proper reward function for the environment. The goal is to learn an optimal policy function to decide the best actions for all states. \par
Particularly, it is required to point out that previous approaches lack the formal definition of heterogeneity. A natural description of the heterogeneity problem is that the action spaces of agents are different, and parameter-sharing among different agents is limited or prohibited. However, such description is not detailed enough for further study. In \cite{bettini2023heterogeneous}, the authors describe and classify the \textit{Physical} and \textit{Behavioral} heterogeneities with natural language instead of mathematical definitions. It is easy for humans to realize that planes and cars are \textit{heterogeneous}. However, it is still necessary to deeply analyze heterogeneity with a formal definition, so that we are able to figure out \textit{what property} is different so that the MASs must be treated differently, and \textit{which type of heterogeneity} do the MASs possess. Based on the definition and classification, we can further quantify and solve the heterogeneity problem. \par
Considering the generation process of a transition tuple, it is concluded that the heterogeneity in MARL mainly occurs in three components of the tuple: Local Reward, Local Observation, and Local Transition. In this paper, we focus on and study the cooperative \textit{Local Transition Heterogeneity} (LTH) MARL problem, in which cooperation happens among different types of agents. When changing the number of ally agents, the ratio of different agent types may also be changed, and thus the optimal cooperating policy is affected. This change increases the diversity and complexity of the LTH problem. \par
A natural solution for LTH is \textit{grouping}. An agent is determined to affiliate a specific group depending on its certain property. Furthermore, an agent keeps to be a permanent member of a group as long as the scenario remains unchanged. The grouping process simplifies and stabilizes the determination of group members and the usage of different group policies, making it easier to add inter-group mechanisms between policies of groups. In addition, grouping helps to maintain a proper structure for parameter-sharing, which helps to improve cooperation through homophily \cite{homophily+dong2021birds}. As a result, it becomes an important problem to choose an appropriate property for grouping in LTH problems.\par
In this paper, we propose GIGM consistency and GHQ algorithm to solve the LTH problem in SMAC environment.
First, in order to leverage the benefit of value-based methods and grouping methods, we need to generalize the Individual-Global Maximum (IGM) consistency \cite{son2019qtran} into grouped situations. Therefore, we conduct the Grouped Individual-Global Maximum (GIGM) consistency and a condition to test whether a grouping method satisfies GIGM.
Second, we propose the Grouped Hybrid Q-Learning (GHQ). Agents are partitioned into groups following the ideal object grouping (IOG) method. Each group has its own isolated network parameters, and the parameters are only shared among group members. A novel \textit{hybrid} structure for value factorization is proposed for optimizing and reducing computation. Furthermore, a variational lower bound of the inter-group mutual information (IGMI) is introduced to increase the correlation between groups for better cooperation. Third, we test GHQ in our new asymmetric heterogeneous maps. Results show that GHQ outperforms other baseline algorithms with higher \textit{WR} and better learning curve, and the cooperate policy between GHQ groups is significantly different against baselines. Main contributions of this paper are:
\begin{itemize}
    \item As far as we know, we are the first to propose the \textit{Local Transition Heterogeneity} (LTH) problem with a formal definition.
    \item We analyze the properties of the LTH problem and design new asymmetric heterogeneous SMAC maps to comprehensively study the LTH problem.
    \item We propose the GIGM consistency and the GHQ algorithm to solve the LTH problem in SMAC.
    \item We run comparison and ablation experiments to prove the effectiveness of the GHQ algorithm.
\end{itemize}
\par
The rest of the content is as follows: we summarize some related works in section \ref{Related Works}; we give the definition of LTH and theoretically analyze it in SMAC in section \ref{Definition and Analysis}; we provide details about the GHQ algorithm in section \ref{Method}; we present detailed environmental and experimental design, and discuss results of our experiments in section \ref{Experiments and Results}; and finally we draw some conclusion in section \ref{Conclusion}.

\begin{table}
\centering
\caption{SMAC Original Maps.}
\label{table-SMAC-maps}
\setlength{\tabcolsep}{1.8mm}{
\begin{tabular}{*{4}{c}}
\toprule
Map Name &  Symmetric  &  Homogeneous  & Difficulty \\
\midrule
1c3s5z & \usym{1F5F8} & \usym{2715} & Easy \\
2c\_vs\_64zg & \usym{2715} & \usym{1F5F8} & Medium\\
2m\_vs\_1z & \usym{2715} & \usym{1F5F8} & Medium\\
2s\_vs\_1sc & \usym{2715} & \usym{1F5F8} & Easy\\
2s3z & \usym{1F5F8} & \usym{2715} & Easy\\
3m & \usym{1F5F8} & \usym{1F5F8} & Easy\\
3s\_vs\_3z & \usym{2715} & \usym{1F5F8} & Easy\\
3s\_vs\_4z & \usym{2715} & \usym{1F5F8} & Medium\\
3s\_vs\_5z & \usym{2715} & \usym{1F5F8} & Medium\\
3s5z & \usym{1F5F8} & \usym{2715} & Easy\\
3s5z\_vs\_3s6z & \usym{2715} & \usym{2715} & Ex-Hard\\
5m\_vs\_6m & \usym{2715} & \usym{1F5F8} & Medium\\
6h\_vs\_8z & \usym{2715} & \usym{1F5F8} & Ex-Hard\\
8m & \usym{1F5F8} & \usym{1F5F8} & Medium\\
8m\_vs\_9m & \usym{2715} & \usym{1F5F8} & Medium\\
10m\_vs\_11m & \usym{2715} & \usym{1F5F8} & Easy\\
25m & \usym{1F5F8} & \usym{1F5F8} & Hard\\
27m\_vs\_30m & \usym{2715} & \usym{1F5F8} & Hard\\
bane\_vs\_bane & \usym{1F5F8} & \usym{2715} & Easy\\
corridor & \usym{2715} & \usym{1F5F8} & Ex-Hard\\
MMM & \usym{1F5F8} & \usym{2715} & Medium\\
MMM2 & \usym{2715} & \usym{2715} & Hard\\
so\_many\_baneling & \usym{2715} & \usym{1F5F8} & Hard\\
\bottomrule
\end{tabular}
}
\end{table}

\section{Related Works}\label{Related Works}
\subsection{Multi-agent Reinforcement Learning}
Following the centralized training with decentralized execution (CTDE) paradigm \cite{ctde+foerster2016learning,ctde+kraemer2016multi,ctde+gupta2017cooperative}, which requests agents not to use state $s$ during execution, recent approaches have achieved impressive results in SMAC environment. The mainstream value-based method is the value factorization method. Its formal objective is to learn a centralized yet factorized joint action-value function $Q_{tot}$ and the factorization structure: $Q_{tot} \to Q_i$, and use them to calculate TD-error and guide the optimization of agent policies:
\begin{equation}
\begin{aligned}
    \mathcal{L}(\theta) &= \mathbb{E}_\mathcal{D} [(r+\gamma \max_{\boldsymbol{a}'} Q_{tot}^{tgt}(\boldsymbol{\tau}', s';\theta^{tgt}) \\
    &-Q_{tot}(\boldsymbol{\tau}, s;\theta))^2] \ ,
\end{aligned}
\end{equation}
where $\mathbb{E}_\mathcal{D}$ means sampling a batch of tuples $(\boldsymbol{\tau}, s, r)$ from replay buffer $\mathcal{D}$ and calculating expectation across the batch. $\boldsymbol{\pi}$ is the action policy, which is commonly the $\epsilon$-greedy policy or \textit{argmax} policy of $Q$ function in value-based algorithms. $Q_{tot}^{tgt}$ is the target function and of $Q_{tot}$. $\theta$ and $\theta^{tgt}$ are the network parameter of $Q_{tot}$ and $Q_{tot}^{tgt}$ respectively. In order to factorize $Q_{tot}$ and use the \textit{argmax} policy of $Q_i$ to select actions, the Individual-Global-Max (IGM) consistency \cite{son2019qtran} is required:
\begin{equation}
\arg \max_{\boldsymbol{a}} Q_{tot}(\boldsymbol{\tau}, s) =
\begin{pmatrix}
  \arg \max_a Q_{1}(\tau_1)\\
  ..., \\
  \arg \max_a Q_k(\tau_k)
\end{pmatrix} \ .
\end{equation}
\par
VDN \cite{vdn+sunehag2017value} represents $Q_{tot}$ as the sum of local $Q_i$ functions. QMIX \cite{qmix+rashid2018} changes the factorization structure from additivity to monotonicity, and the fine-tuned version of QMIX has been proved to be one of the best algorithms on the original SMAC maps \cite{riit+hu2021rethinking}. Based on these two fundamental algorithms, QTRAN \cite{son2019qtran}, WQMIX \cite{wqmix+rashid2020weighted}, Qatten \cite{yang2020qatten}, and QPLEX \cite{wang2020qplex} improve performance with modified value factorizing mechanism. Qauxi \cite{QMIX-auxi+liang2022qauxi} introduces auxiliary tasks to generate meta-experience for transfer learning. CDCR \cite{APPI-SMAC-ge2022enhancing} calculates Cognition Differences between agents with the attention mechanism and learns Consistent Representation of agents' hidden states for enhancing cooperation. Trans\_mix \cite{APPI-SMAC-wang2022transform} uses a transform network to solve the misalignment of partially observatory value. BRGR \cite{APPI-SMAC-he2023brgr} uses Bidirectional Real-time Gain Representation to learn overall information representation and neighbor information representation, and combines them with other value-based algorithms for better cooperation.\par
Heterogeneous MARL has been considered as a special case of homogeneous MARL and can be handled with individual policy networks. HAPPO \cite{happo+kuba2021trust}, in which the $H$ stands for heterogeneous, lacks specific analysis and sufficient experiments for heterogeneity. In other field of MAS, \cite{robotics+yang2021can} uses Relative Needs Entropy (RNE) to build a trust model to improve cooperation in heterogeneous multi-robot grouping task, and \cite{robotics+hartmann2021long} contributes a novel method for the heterogeneous multi-robot assembly planning.

\subsection{Grouping Method}\label{Related Work-Grouping}
Grouping is a natural idea and solution for complex or large-scale problems and is widely used in many research of optimization or machine learning. In SMAC environment, THGC \cite{APPI-SMAC-jiang2021multi} divides agents into different groups based on their different ``types'' for knowledge sharing and group communication. However, it is necessary to formally define and describe the difference between agent types in a universal way across different environments. In this paper, we introduce some auxiliary definitions for describing our grouping method. \par
\cite{grouping+liu2020multi} uses a channel grouping algorithm to cluster different sub-regions of pictures for vehicle Re-ID. \cite{grouping+li2021differential} introduces a ranking-based grouping method to improve multi-population-based differential evolution algorithm. \cite{grouping+cheng2021hybrid} proposes a grouping attraction model, which can significantly reduce the number of attractions and fitness comparisons in the firefly algorithm. \cite{grouping+li2021deep} modifies the Transformer encoder by properly organizing encoder layers into multiple groups, and connects these groups via a grouping skip connection mechanism. \cite{grouping+rotman2020learnable} enhances the Optimal Sequential Grouping (OSG) to solve the video scene detection problem. \cite{grouping+ling2022fedentropy} proposes FedEntropy for better dynamic device grouping in federated learning. \cite{grouping+hou2022enhanced} introduces an enhanced decentralized autonomous aerial swarm system with group planning. \cite{grouping+al2021self} designs a self-organizing MAS for distributed voltage regulation in the smart power grid.

\subsection{Mutual Information}
Computing the variational bound of mutual information (MI) has been proven to enhance cooperation in MARL. MAVEN \cite{mahajan2019maven} maximizes a variational lower bound of the MI between the latent variable $z$ and the agent-specific Boltzmann policy $\sigma(\tau)$ to encourage exploration of the algorithm. ROMA \cite{wang2020roma} computes two MI-related losses to learn identifiable and specialized role policies. PMIC \cite{li2022pmic} maintains positive and negative trajectory memories to compute the upper bound and lower bound of the MI between global state $s$ and joint action $\boldsymbol{a}$. MAIC \cite{yuan2022+maic} maximizes the MI between the trajectory of agent $i$ and the ID of another agent $j$ for teammate modeling and communication. CDS \cite{li2021+CDS} maximizes the MI between the trajectory $\tau_i$ of agent $i$ and its own agent ID to maintain diverse individual local $Q$ functions.

\section{Local Transition Heterogeneity}\label{Definition and Analysis}
In this section, our goal is to give a formal definition of the \textit{Local Transition Heterogeneity} (LTH) problem and analyze its existence in SMAC. We first present fundamental concepts and definitions in \ref{Preliminaries}. Next, we define auxiliary concepts and the Local Transition Function (LTF) for the formal definition of LTH in \ref{LTF}. These definitions isolate one specific agent $i$ into an ideal scenario. Therefore, we can study the properties of agent $i$ affecting the LTH problem. And then, in \ref{LTH}, we define the LTH problem and show the advantage of our definition. Finally, we conclude two properties for proving the existence of the LTH problem, and analyze the existence of LTH in SMAC in \ref{LTH in SMAC}.

\subsection{Preliminaries}\label{Preliminaries}
In this paper, we study the cooperative MARL problems that can be modeled as the decentralized partially observable Markov decision process (Dec-POMDP) \cite{DEC-POMDP+oliehoek2016concise}. The problem is described with a tuple $G=\left\langle S, \boldsymbol{A}, \boldsymbol{P}, R, \boldsymbol{\Omega}, O; \gamma, K, T \right\rangle$. $s \in S$ denotes the true state of environment with complete information, $K= \{ 1, ..., k  \} $ denotes the finite set of $k$ agents, and $\gamma \in [0,1)$ is the discount factor. At each time-step $t \le T$, agent $i \in K$ receives an individual partial observation $o_i^t$ and chooses an action $a_i^t \in A_i$ from local action set $A_i$, with the local action-dim $|A_i|$. Actions of all agents form a joint action $\boldsymbol{a}^t = (a_1^t, ..., a_k^t) \in \boldsymbol{A} = (A_1, ..., A_k)$. The environment receives a joint action $\boldsymbol{a}^t$ and returns a next-state $s^{t+1}$ according to the joint transition function $\boldsymbol{P} (s^{t+1}|s^{t}, \boldsymbol{a}^t)$, and a reward $r^t = R(s, \boldsymbol{a}^t)$ shared by all agents. The joint observation $\boldsymbol{o}^t = (o_1^t, ..., o_k^t) \in \boldsymbol{\Omega}$ is generated according to the observation function $O^t (s^t, i)$. Observation-action trajectory history $\tau^t = \cup_1^t \{ (o^{t-1}, a^{t-1})\}$ ($t \ge 1; \tau^0 = o^0$) is the summary of partial transition tuples before $t$. Specifically, $\boldsymbol{\tau}_i = \boldsymbol{\tau}_i^T$ indicates the overall trajectory of agent $i$ through all time-steps $t \le T$. Replay buffer $\mathcal{D} = \cup (\boldsymbol{\tau}, s, r)$ stores all data for batch sampling. Network parameters are notated by $\theta$ and $\psi$.

\subsection{Auxiliary Definitions and the Local Transition Function}\label{LTF}
Apart from the joint transition function $\boldsymbol{P} (s^{t+1}|s^{t}, \boldsymbol{a}^t)$, we need to define the Local Transition Function (LTF) $P_i (s^{t+1}|s^{t}, a_i^t)$ for the definition of LTH problem. Several auxiliary definitions are given for better demonstration and analysis of LTF and LTH. \par
First, we partition the actions of an agent $A_i$ into 3 different types: common actions $A_{com}$, which only affect agent $i$ itself, \textit{e.g.} moving, scanning and transforming; interactive actions $A_{act}$, which are interacting with other agents, \textit{e.g.} attacking, guiding and delivering; and mixing actions $A_{mix}$, which affect both itself and others, \textit{e.g.} a predator moving close to a prey for automatic predating. Usually, $A_{mix}$ can be divided into the combination of $A_{com}$ and $A_{act}$, \textit{e.g.} the $A_{mix}$ automatic predating can be divided into $A_{com}$ moving and $A_{act}$ predating. For terminological simplicity, we divide $A_{mix}$ into the combination of $A_{act}$ and $A_{com}$ by default, and focus on the latter two types of actions. \par
Second, we introduce the joint available-action-mask matrix $\boldsymbol{AM} (s^t)$ and the local available-action-mask vector $AM_i (s^t, i)$, which are common components in many MARL environments. $\boldsymbol{AM} (s^t)$ is a binary matrix with dimensions being $|A_i| \times K$, indicating the available-actions of all agents at the state $s^t$. $AM_i (s^t, i)$ is the column vector of $\boldsymbol{AM} (s^t)$, indicating the mask vector of a certain agent. Element 1 (true) at $(a_i^t, i)$ of $\boldsymbol{AM} (s^t)$ means that agent $i$ can take action $a_i^t$ at $s^t$, and vice versa. \par
Finally, we define the Ideal Object $(IO)$ and the Ideal Condition $(IC)$, and then define the LTF, $P_i (s^{t+1}|s^{t}, a_i^t)$. \par

\begin{definition}
Ideal Object (IO) and Ideal Condition (IC): The Ideal Object $(IO_i)$ of agent $i$ is an action object that is available for any $A_{act}$ of agent $i$ to be applied on. The Ideal Condition $(IC_i)$ of agent $i$ is the environmental condition that maintains the local available-action-mask function $AM_i (s^t, i)$ being all true for any state $s^t$ and any action $a_i^t$ applied on $IO_i$.
\end{definition}

\begin{definition}
Local Transition Function (LTF): For agent $i$ with its $IO_i$ and $IC_i$, the Local Transition Function (LTF) $P_i (s^{t+1}|s^t, a_i^t)$ is the probability distribution of next-state $s^{t+1}$ conditioned by state $s^t$ and action $a_i^t$. The action $a_i^t$ is applied on $IO_i$ under $IC_i$.
\end{definition}

\subsection{Definition of Local Transition Heterogeneity}\label{LTH}
In general, the \textit{Local Transition Heterogeneity} (LTH) means that agents cannot reach the same next-state $s^{t+1}$ from the same state $s^{t}$, no matter what policies they are using. A formal definition is given below. \par

\begin{definition}
Local Transition Heterogeneity (LTH): Let there be two agents $i,j \in K$. Their policies are $\pi_i(a_i|s)$ and $\pi_j(a_j|s)$, and their LTFs are $P_i (s^{t+1}|s^{t}, a_i^t)$ and $P_j (s^{t+1}|s^{t}, a_j^t)$. A certain state $s^t$, which simultaneously fulfills $IC_i$ and $IC_j$, is the starting state. The sets of next-states $\{s^{t+1}_i | s^{t}, \pi_i, P_i\}$ and $\{s^{t+1}_j | s^{t}, \pi_j, P_j\}$ are generated by $\pi_i$ and $\pi_j$ individually executed on $s^{t}$ towards their corresponding $IO_i$ and $IO_j$. If the intersection of the two sets of next-states is empty for all available policies, then the MARL problem has LTH:
\begin{equation}
\begin{aligned}
    &\exists \ i,j \in K,\ \forall \ \pi_i, \pi_j,\\
    &\{s^{t+1}_i | s^{t}, \pi_i, P_i\} \cap \{s^{t+1}_j | s^{t}, \pi_j, P_j\} = \emptyset \ .
\end{aligned}
\label{LTH-definition}
\end{equation}
\end{definition}
\par

For example, in a MAS consisting of UAV (Unmanned Aerial Vehicle) and UGV (Unmanned Ground Vehicle), we suppose that UAV and UGV carry different mission cargo, so their $A_{act}$ are different. Their moving speed and moving dimension (2-D and 3-D) are also different, so their $A_{com}$ are different. From the same starting state $s^{t}$, UAV and UGV cannot reach the same next-state $s^{t+1}$ because their action spaces are completely different. Therefore, the existence of LTH problem in such MAS is clear. \par
An advantage of our definition is the reliability of presenting heterogeneity. We define the LTH problem under the restriction of $IO$ and $IC$. Our core motivation is to ensure that the local available-action-mask vector $AM_i (s^t, i)$ remains all true, because $AM_i (s^t, i)$ can influence the behavior of agents and thus affect the existence of LTH. For instance, if all enemies choose the policy ``attack and eliminate agent $i$ at the $1^{st}$ time-step'', then the $AM_i$ would be only available for the ``dead-action'', since agent $i$ is always dead from the $1^{st}$ time-step. Therefore, it is impossible for agent $i$ to present LTH. Similarly, ally agents' actions and policies can also affect the $AM_i$ and lead to the same result. In conclusion, our definition avoids unexpected influence from enemies or allies towards the $AM_i$, and is capable of presenting LTH reliably. \par

\begin{table}
\centering
\caption{Unit Information in SMAC. DPS stands for ``damage-per-game-second''. For Medivac, the number of DPS indicates the healing-per-game-second (HPS).}
\label{table-SMAC-units}
\setlength{\tabcolsep}{1.0mm}{
\begin{tabular}{*{7}{c}}
\toprule
Unit & Health- & Shot- & \multirow{2}*{DPS} & \multirow{2}*{Speed} & Unit & Flying\\
Name &  point  & range &  &  & Type & Unit\\
\midrule
Marine & 45 & 5 & 6.97 & 2.25 & $U_{atk}$ & \usym{2715}\\
Medivac & 150 & 4 & 9.00 & 2.75 & $U_{spt}$ & \usym{1F5F8}\\
Marauder & 125 & 6 & 6.67 & 2.25 & $U_{atk}$ & \usym{2715}\\
\bottomrule
\end{tabular}
}
\end{table}

\subsection{Existence of LTH in SMAC}\label{LTH in SMAC}
The original definition formula (\ref{LTH-definition}) is inconvenient to judge whether an environment has LTH. We further conclude that the difference of $IO$ or LTF can determine the existence of LTH. First, different $IO$ leads to qualitative LTH. For example, in a UAV-UGV system with different mission cargo, the $IO$ of a UAV is defined to be another UAV while the $IO$ of a UGV is defined to be another UGV. Their objects and functionalities of $A_{act}$ are different, leading to LTH. Generally, different interactive action-dim $|A_{act}|$ is sufficient to prove the difference of $IO$, and can also be used to prove the existence of LTH. Second, different LTFs lead to quantitative LTH. For example, in a UAV-UGV system with the same mission cargo, their moving speeds are still different. Typically, UAVs fly faster in the air than UGVs move on the ground. The difference of the dynamics of $A_{com}$ or $A_{act}$ leads to different LTF, and thus LTH occurs. \par
In SMAC, there are two agent types, supporting units $U_{spt}$ and attacking units $U_{atk}$. $U_{spt}$ can only affect allies while $U_{atk}$ can only affect enemies. For example, Medivac is a $U_{spt}$ who can only heal allies, while Marine is a $U_{atk}$ who can only attack enemies (see Table \ref{table-SMAC-units}). In SMAC, $A_{com}$ are moving and stopping, available for all living agents at any state $s$ and any time-step $t$. The common action-dim $|A_{com}|$ also remains identical among all agent types. $A_{act}$ are attacking or healing. A certain agent type can only attack enemies or heal allies. Therefore, $|A_{act}|$ should be different between different agent types. \par
First, the $IO$ of $U_{atk}$ and $U_{spt}$ are different, leading to qualitative LTH. For $U_{atk}$, its $IO$ is an enemy, and its $|A_{act}|$ is also the total number of \textit{enemies}. However, for $U_{spt}$, its $IO$ is an ally, so its $|A_{act}|$ is the total number of \textit{allies}. Second, the moving \textit{speed}, \textit{shot-range} and \textit{damage-per-gaming-second} (DPS) are different between different types of agents (see Table \ref{table-SMAC-units}), indicating the existence of quantitative LTH. In conclusion, the existence of LTH in SMAC is clarified, and further analysis and study of LTH are therefore required.\par

\section{Method}\label{Method}
\subsection{Grouped Individual-Global-Max Consistency}
As is shown in section \ref{LTH}, LTH does not change the reward function $R(s, \boldsymbol{a})$ or the available-action-mask. Therefore, any available joint action $\boldsymbol{a}$ is rewarded the same as it in homogeneous scenarios, and the optimal joint action $\boldsymbol{a}^*$ is not affected. As a result, the IGM consistency in LTH still holds and we can further generalize the consistency to a ``grouped'' situation for solving LTH problems with grouping value factorization.\par

\begin{definition}
Grouped IGM Consistency (GIGM): Let there be $U= \left \{ 1, ..., u \right \}, \ (u<k)$ agent groups in total. An agent group $\mathcal{G}_m\ (m \in U)$ consists of agents arbitrarily pre-defined. If the \textit{argmax} operation performed on the joint function $Q_{tot}$ yields the same result as a set of individual \textit{argmax} operations performed on all group functions $Q_{\mathcal{G}_m}\ (m \in U)$; and the \textit{argmax} operation performed on each group function $Q_{\mathcal{G}_m}$ yields the same result as a set of individual \textit{argmax} operations performed on the agent functions $Q_i\ (i \in \mathcal{G}_m)$, then GIGM holds true:

\begin{equation}
\begin{aligned}
&\arg \max_{\boldsymbol{a}} Q_{tot}(\boldsymbol{\tau}, s) \\
&=
\begin{pmatrix}
  \arg \max_{\boldsymbol{a}} Q_{\mathcal{G}_1}(\boldsymbol{\tau}_{\mathcal{G}_1}, s)\\
  ..., \\
  \arg \max_{\boldsymbol{a}} Q_{\mathcal{G}_u}(\boldsymbol{\tau}_{\mathcal{G}_u}, s)
\end{pmatrix} \\
&=
\begin{pmatrix}
  \arg \max_a Q_1(\tau_1)\\
  ..., \\
  \arg \max_a Q_k(\tau_k)
\end{pmatrix},
\end{aligned}
\end{equation}

\begin{equation}
\begin{aligned}
&\arg \max_{\boldsymbol{a}} Q_{\mathcal{G}_m}(\boldsymbol{\tau}_{\mathcal{G}_m}, s) \\
&=
\begin{pmatrix}
  \arg \max_a Q_i(\tau_i)\\
  i \in \mathcal{G}_m
\end{pmatrix},
\end{aligned}
\end{equation}
where $\boldsymbol{\tau}_{\mathcal{G}_m} = \cup_{i \in \mathcal{G}_m}\{\tau_i\}$ is the group trajectory of $\mathcal{G}_m$, $\boldsymbol{\tau} = \cup_{i \in K}\{\tau_i\}$ is the global joint trajectory of all agents.
\end{definition}
\par

Furthermore, We conclude a theorem sufficient to prove GIGM: \par
\begin{theorem}
Joint Trajectory Condition (JTC): GIGM holds true if the following two conditions are simultaneously satisfied: \par
(i)\ The global joint trajectory is equivalent to the union of all group trajectories.
\begin{equation}
    \boldsymbol{\tau} = \cup_{i \in K}\ \{\tau_i\} = \cup_{m \in U}\ \{\boldsymbol{\tau}_{\mathcal{G}_m}\}\ .
\end{equation}
\par
(ii)\ The intersection of all group trajectories is empty.
\begin{equation}
    \cap_{m \in U}\ \{\boldsymbol{\tau}_{\mathcal{G}_m}\} = \emptyset \ .
\end{equation}
\end{theorem}
\par

The first condition guarantees the transitivity of \textit{argmax} operations performed on $Q$ functions. The second condition guarantees the coexistence of \textit{argmax} operations on all $Q_{\mathcal{G}_m}$. The two conditions jointly guarantee the equivalence of \textit{argmax} operations on all group $Q$ and agent $Q$ functions:
\begin{equation}
    \arg \max_{\boldsymbol{a}} \ Q_{\mathcal{G}_m}\ (m \in U)\ =\ \arg \max_a \ Q_i\ (i \in K) \ .
\end{equation}

\subsection{Ideal Object Grouping}
In order to utilize GIGM to solve the LTH problem, we propose Ideal Object Grouping (IOG), which means partitioning agents into different groups by their different ideal objects $IO$. As is mentioned in section \ref{Related Work-Grouping}, we need to formally define and describe `` what is the meaning of different types of agents'' in a universal way. And we point out that the difference in $IO$ is equivalent to the difference of agent types, because fundamentally these differences are all about the difference in agent action space $\boldsymbol{A}$. This is the exact functionality and property describing the heterogeneous agents. In general, our goal is to acquire a grouping function $g(i, {\mathcal{G}_m})\ (i\in K,\ m\in U)$ for agent $i$ and group $\mathcal{G}_m$:
\begin{equation}
    g(i, {\mathcal{G}_m}) = \begin{cases}
1  & \text{ if } i \in {\mathcal{G}_m} \\
0  & \text{ else }
\end{cases} \ .
\end{equation}
\par
Each agent group $\mathcal{G}_m$ consists of agents with the same $IO_{\mathcal{G}_m}$ and the same interactive action-dim $|A_{act-\mathcal{G}_m}|$. Only one universal agent network is kept for one group, which significantly reduces the number of agent networks from $K$ to $U$. Parameter-sharing is only allowed between agents within the same group. Maintaining a proper parameter-sharing structure not only avoids redundant computing resources for individual agent networks, but can also increase in-group cooperating via homophily \cite{homophily+dong2021birds}. \par
Moreover, IOG is a mapping function from agents to groups $g(i, {\mathcal{G}_m}): (K \to U)$. The $|A_{act}|$ of each agent must be assigned during the initialization of SMAC. As a result, one specific agent $i$ can only be assigned to the certain group with $IO_i = IO_{\mathcal{G}_m}$ and $|A_{act_i}| = |A_{\mathcal{G}_m}|$. Therefore, JTC is satisfied and GIGM holds true, indicating that IOG is an appropriate grouping method for value factorization.

\subsection{Inter-Group Mutual Information Loss}\label{IGMI-loss}
\begin{figure*}
\centering
\includegraphics[width=0.99\textwidth]{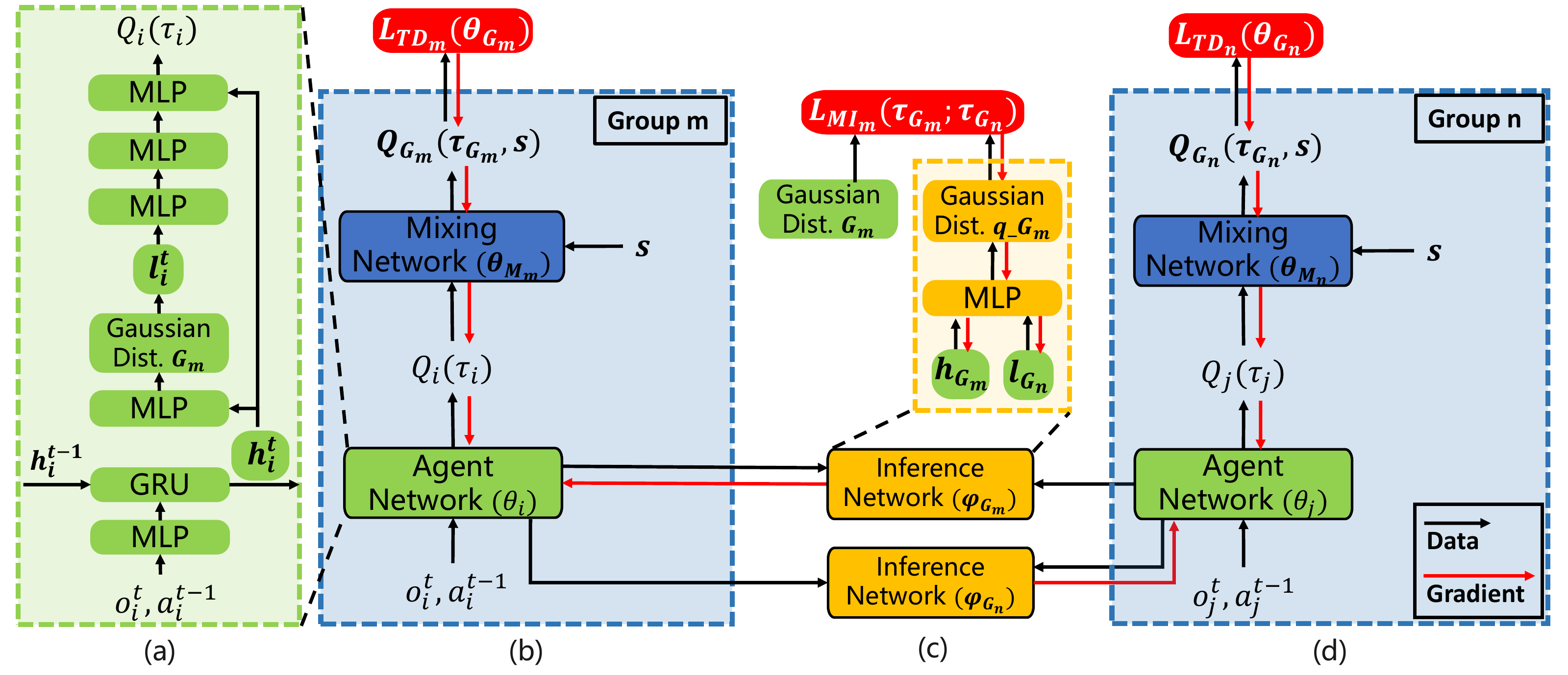}
\caption{An overall framework of GHQ. $\boldsymbol{\theta}_{\mathcal{G}_m}$ of group $m$ consists of three parts: agent network $\theta_i$, mixing network $\theta_{M_m}$ and inference network $\boldsymbol{\psi}_{\mathcal{G}_m}$. Detailed data-stream for training and executing are shown in Fig. \ref{fig-data_stream}. In (a), $\theta_i$ takes $o_i^t$ and $a_i^{t-1}$ as input. It generates $Q_i$ for choosing actions, and $l_{\mathcal{G}_m}$ and $h_{\mathcal{G}_m}$ for calculating $q_{\mathcal{G}_m}$. In (b), $\theta_{M_m}$ takes $Q$ and $s$ for calculating TD loss $\mathcal{L}_{TD_m}$ with hybrid factorization. In (c), $\boldsymbol{\psi}_{\mathcal{G}_m}$ takes $l_{\mathcal{G}_m}$, $h_{\mathcal{G}_m}$ and $l_{\mathcal{G}_n}$ for calculating IGMI loss $\mathcal{L}_{MI_m}$.}
\label{fig-framework}
\end{figure*}

In order to enhance inter-group cooperation and correlation, we maximize the Inter-Group Mutual Information (IGMI) between trajectories of different groups ${\boldsymbol{\tau}_{\mathcal{G}_m}}$ and ${\boldsymbol{\tau}_{\mathcal{G}_n}}$, written as $I(\boldsymbol{\tau}_{\mathcal{G}_m}; \boldsymbol{\tau}_{\mathcal{G}_n})$. For encoding trajectories, a common implementation is to use the hidden states of \textit{gated recurrent unit} (GRU) \cite{cho2014+GRU} $h_{\mathcal{G}_m}$ and $h_{\mathcal{G}_n}$. While GRU takes $\boldsymbol{o}_{\mathcal{G}_m}^t$ and $\boldsymbol{a}_{\mathcal{G}_m}^t$ recursively for all $t \le T$, we assume that $h_{\mathcal{G}_m}$ is capable of encoding and representing $\boldsymbol{\tau}_{\mathcal{G}_m}$. After encoding, because the mutual information can only be calculated between two distributions, we add a Gaussian distribution layer in the agent network of every group, marked as $l_{\mathcal{G}_m}$ and $l_{\mathcal{G}_n}$. Therefore, calculating $I(\boldsymbol{\tau}_{\mathcal{G}_m}; \boldsymbol{\tau}_{\mathcal{G}_n})$ can be converted into calculating $I(l_{\mathcal{G}_m}; l_{\mathcal{G}_n} | h_{\mathcal{G}_m}, h_{\mathcal{G}_n})$. Detailed agent network structure is illustrated in Fig. \ref{fig-framework}(a).
\begin{equation}
\begin{aligned}
    &h_{\mathcal{G}_m} = GRU(\boldsymbol{\tau}_{\mathcal{G}_m}) = GRU(\cup_{i \in \mathcal{G}_m}\{\tau_i\})\ ,\\
    &l_{\mathcal{G}_m} = Gaussian(h_{\mathcal{G}_m})\ ,\\
    &I(\boldsymbol{\tau}_{\mathcal{G}_m}; \boldsymbol{\tau}_{\mathcal{G}_n})=I(l_{\mathcal{G}_m}; l_{\mathcal{G}_n} | h_{\mathcal{G}_m}, h_{\mathcal{G}_n})\ .
\end{aligned}
\end{equation}
\par
We further conduct a lower bound of $I(\boldsymbol{\tau}_{\mathcal{G}_m}; \boldsymbol{\tau}_{\mathcal{G}_n})$ for easier calculation:
\begin{equation}
\begin{aligned}
    &I(\boldsymbol{\tau}_{\mathcal{G}_m}; \boldsymbol{\tau}_{\mathcal{G}_n})\\
    &=I(l_{\mathcal{G}_m}; l_{\mathcal{G}_n} | h_{\mathcal{G}_m}, h_{\mathcal{G}_n})\\
    &=\mathbb{E}_{l_{\mathcal{G}_m}, l_{\mathcal{G}_n}, h_{\mathcal{G}_m}, h_{\mathcal{G}_n}}[log\frac{p(l_{\mathcal{G}_m}|l_{\mathcal{G}_n}, h_{\mathcal{G}_m}, h_{\mathcal{G}_n})}{p(l_{\mathcal{G}_m})}]\\
    &=\mathbb{E}_{l_{\mathcal{G}_m}, l_{\mathcal{G}_n}, h_{\mathcal{G}_m}, h_{\mathcal{G}_n}}[log\ p(l_{\mathcal{G}_m}|l_{\mathcal{G}_n}, h_{\mathcal{G}_m}, h_{\mathcal{G}_n}) \\
    &- log\ p(l_{\mathcal{G}_m}) + log\ q_{\mathcal{G}_m}(l_{\mathcal{G}_m}|l_{\mathcal{G}_n}, h_{\mathcal{G}_m}, h_{\mathcal{G}_n})\\
    &- log\ q_{\mathcal{G}_m}(l_{\mathcal{G}_m}|l_{\mathcal{G}_n}, h_{\mathcal{G}_m}, h_{\mathcal{G}_n})]\\
    &=\mathbb{E}_{l_{\mathcal{G}_m}, l_{\mathcal{G}_n}, h_{\mathcal{G}_m}, h_{\mathcal{G}_n}}[log\frac{q_{\mathcal{G}_m}(l_{\mathcal{G}_m}|l_{\mathcal{G}_n}, h_{\mathcal{G}_m}, h_{\mathcal{G}_n})}{p(l_{\mathcal{G}_m})}\\
    &+\alpha * D_{KL}(p(l_{\mathcal{G}_m}|l_{\mathcal{G}_n}, h_{\mathcal{G}_m}, h_{\mathcal{G}_n}) \\
    &|| q_{\mathcal{G}_m}(l_{\mathcal{G}_m}|l_{\mathcal{G}_n}, h_{\mathcal{G}_m}, h_{\mathcal{G}_n}))]\ ,
\end{aligned}
\end{equation}
where $\alpha = \frac{p(l_{\mathcal{G}_m})}{p(l_{\mathcal{G}_m}|l_{\mathcal{G}_n}, h_{\mathcal{G}_m}, h_{\mathcal{G}_n})}$ is always non-negative, and $D_{KL}$ is the KL-divergence being also non-negative. $q_{\mathcal{G}_m}$ is an inference distribution of group $\mathcal{G}_m$ with parameter $\psi_{\mathcal{G}_m}$, and is independent from $h_{\mathcal{G}_n}$. To keep this independence, a mixed input of different groups is forbidden. Therefore, we keep individual inference networks for each group. Finally, we have:
\begin{equation}
\begin{aligned}
    &I(\boldsymbol{\tau}_{\mathcal{G}_m}; \boldsymbol{\tau}_{\mathcal{G}_n})\\
    &\ge \mathbb{E}_{l_{\mathcal{G}_m}, l_{\mathcal{G}_n}, h_{\mathcal{G}_m}, h_{\mathcal{G}_n}}[log\frac{q_{\mathcal{G}_m}(l_{\mathcal{G}_m}|l_{\mathcal{G}_n}, h_{\mathcal{G}_m}, h_{\mathcal{G}_n})}{p(l_{\mathcal{G}_m})}]\\
    &=\mathbb{E}_{l_{\mathcal{G}_m}, l_{\mathcal{G}_n}, h_{\mathcal{G}_m}}[log\ q_{\mathcal{G}_m}(l_{\mathcal{G}_m}|l_{\mathcal{G}_n}, h_{\mathcal{G}_m})\\
    &- log\ p(l_{\mathcal{G}_m})]\\
    &=\mathbb{E}_{l_{\mathcal{G}_n}, h_{\mathcal{G}_m}}[-CE(p(l_{\mathcal{G}_m}) || q_{\mathcal{G}_m}(l_{\mathcal{G}_m}|l_{\mathcal{G}_n}, h_{\mathcal{G}_m}))\\
    &+ \mathcal{H}(p(l_{\mathcal{G}_m}))]\\
    &=-\mathbb{E}_{l_{\mathcal{G}_n}, h_{\mathcal{G}_m}}[D_{KL}(p(l_{\mathcal{G}_m}) || q_{\mathcal{G}_m}(l_{\mathcal{G}_m}|l_{\mathcal{G}_n}, h_{\mathcal{G}_m}))] \ .
\end{aligned}
\end{equation}
\par
In order to maximize the IGMI, the loss is written as:
\begin{equation}
\begin{aligned}
    &\mathcal{L}_{MI_m}(\tau_{\mathcal{G}_m}; \tau_{\mathcal{G}_n} | \psi_{\mathcal{G}_m}) \\
    &=\mathbb{E}_\mathcal{D} [D_{KL}(p(l_{\mathcal{G}_m})||q_{\mathcal{G}_m}(l_{\mathcal{G}_m} | l_{\mathcal{G}_n}, h_{\mathcal{G}_m}))]\ .
    \label{equa-Loss-MI}
\end{aligned}
\end{equation}

\begin{algorithm}
\caption{GHQ}
\label{algo-GHQ}
\textbf{Input}: Learning rate $\alpha$, loss weights $\lambda_{TD}$ and $\lambda_{MI}$, number of groups $U$, number of agents $K$, number of units in each group $|\mathcal{G}_m|$, max total steps $T^{TOT}$, max steps per episode $T^{EP}$, batch size $B$.\\
\textbf{Initialize}: Network parameters $\boldsymbol{\theta}=\{\boldsymbol{\theta}_{\mathcal{G}_{m}}\}_1^U$ and $\boldsymbol{\psi}=\{\boldsymbol{\psi}_{\mathcal{G}_{m}}\}_1^U$, replay buffer $\mathcal{D} = \{\}$, total step $t^{TOT}=0$.
\begin{algorithmic}[1] 
\While{ $t^{TOT}\le T^{TOT}$}
\State Set episode step $t=0$, receive initial state and observation $(s^0, o^0)$ from environment.
\While{$t \le T^{EP}$ and \textbf{not} $terminated$}
\State Choose joint action $\boldsymbol{a}^t =\{a_i^{t}\}_1^K$ with agent networks $\{\theta_i\}_1^K$.
\State Receive $r^{t}$, \textit{is\_terminated} and $(s^{t+1}, \boldsymbol{o}^{t+1})$ from environment using $\boldsymbol{a}^t$.
\State Collect a transition tuple at $t$ and update the replay buffer $\mathcal{D} = \mathcal{D} \cup \{(\boldsymbol{\tau}^{t}, s^{t}, r^{t})\}$.
\State $t=t+1$.
\EndWhile
\State Sample a random batch of $B$ episodes from $\mathcal{D}$.
\For{$m=0, ..., U-1$}
\State Calculate $\mathcal{L}_{MI_m}$ following (\ref{equa-Loss-MI}).
\State Calculate $\mathcal{L}_{TD_m}$ following (\ref{equa-Loss-TD}).
\EndFor
\State Calculate total loss $\mathcal{L}_{GHQ}$ following (\ref{equa-Loss-GHQ}) and update network parameters $\boldsymbol{\theta}$ and $\boldsymbol{\psi}$.
\State $t^{TOT}=t^{TOT}+t$.
\EndWhile
\end{algorithmic}
\end{algorithm}

\subsection{Grouped Hybrid Q-Learning}
An ordinary idea to calculate $Q_{\mathcal{G}_m}$ and $Q_i$ is to design factorization structures for $Q_{tot} \to Q_{\mathcal{G}_m}$ and $Q_{\mathcal{G}_m} \to Q_i$. Let $\mathcal{C}_{\mathcal{G}_m}$ and $\mathcal{C}_i$ be the two factor function. Like IGM \cite{rashid2020+qmix-jmlr}, the monotonicity constraint is also sufficient for GIGM. Therefore, $\mathcal{C}_{\mathcal{G}_m}$ and $\mathcal{C}_i$ can be written as:
\begin{equation}
    \frac{\partial Q_{tot}(\boldsymbol{\tau}, s)}{\partial Q_{\mathcal{G}_m}(\boldsymbol{\tau}_{\mathcal{G}_m}, s)} = \mathcal{C}_{\mathcal{G}_m} \ge 0,\ m \in U \ ,
    \label{equa-frac-Qtot}
\end{equation}
\begin{equation}
    \frac{\partial Q_{\mathcal{G}_m}(\boldsymbol{\tau}_{\mathcal{G}_m}, s)}{\partial Q_i(\tau_i)} = \mathcal{C}_i \ge 0,\ i \in \mathcal{G}_m \ .
    \label{equa-frac-Qgi}
\end{equation}
\par
Our key insight is that trivially calculating $\mathcal{C}_{\mathcal{G}_m}$ and $Q_{\mathcal{G}_m}$ is unnecessary. Instead of hierarchical factorization, we imply \textit{independent Q-Learning} (IQL) \cite{tan1993+IQL} for $\mathcal{C}_{\mathcal{G}_m}$, which is called the \textit{hybrid} factorization. This method makes $Q_{\mathcal{G}_m}$ become an action-value function instead of a utility function \cite{guestrin2001+utility,rashid2020+qmix-jmlr}, and $\mathcal{C}_{\mathcal{G}_m}$ become a positive constant. As a result, the TD loss of group $\mathcal{G}_m$ is written as:
\begin{equation}
\begin{aligned}
    &\mathcal{L}_{TD_m}(\boldsymbol{\theta}_{\mathcal{G}_m})\ =\
    \mathbb{E}_\mathcal{D} [(y^{\mathcal{G}_m}-Q_{\mathcal{G}_m}(\boldsymbol{\tau}_{\mathcal{G}_m},s;\boldsymbol{\theta}_{\mathcal{G}_m}))^2]\ ,\\
    &y^{\mathcal{G}_m}\ =\ r+\gamma \ \max_{\boldsymbol{a}'} Q_{\mathcal{G}_m}^{tgt}(\boldsymbol{\tau}_{\mathcal{G}_m}',s';\boldsymbol{\theta}^{tgt}_{\mathcal{G}_m})\ ,
    \label{equa-Loss-TD}
\end{aligned}
\end{equation}
where $y^{\mathcal{G}_m}$ is the TD-target of $Q_{\mathcal{G}_m}$, $Q_{\mathcal{G}_m}^{tgt}$ is the target Q function of $Q_{\mathcal{G}_m}$, and $\boldsymbol{\theta}^{tgt}_{\mathcal{G}_m}$ and $\boldsymbol{\theta}_{\mathcal{G}_m}$ are network parameters of $Q_{\mathcal{G}_m}^{tgt}$ and $Q_{\mathcal{G}_m}$, separately. Group network $\boldsymbol{\theta}_{\mathcal{G}_m}$ consists of two parts, agent network $\theta_i$ and mixing network $\theta_{M_m}$. Their losses are calculated with backward propagation following (\ref{equa-frac-Qgi}):
\begin{equation}
    \frac{\partial \mathcal{L}_{TD_m}(\boldsymbol{\theta}_{\mathcal{G}_m})}{\partial \theta_i} = \frac{\partial \mathcal{L}_{TD_m}(\boldsymbol{\theta}_{\mathcal{G}_m})}{\partial \theta_{M_m}} \cdot \frac{\partial \theta_{M_m}}{\partial \theta_i}\ .
\end{equation}
\par
GIGM and the input of state information keep different $Q_{\mathcal{G}_m}$ in relevance, and IGMI further enhances the correlation. Even though IQL methods suffer from non-stationary problems \cite{foerster2017+iql-Stabilising}, GHQ overcomes this disadvantage and achieves impressive results. The hybrid factorization avoids the calculation hierarchical factorization function. Although the IQL value of $Q_{\mathcal{G}_m}$ following (\ref{equa-Loss-TD}) does not equal the factorized value of $Q_{\mathcal{G}_m}$ following (\ref{equa-frac-Qtot}), the monotonicity of factorization and GIGM still hold. As a result, the optimal policy of GHQ converges to the same optimal policy provided by the fully factorized structure.

\begin{figure*}
\centering
\includegraphics[width=0.99\textwidth]{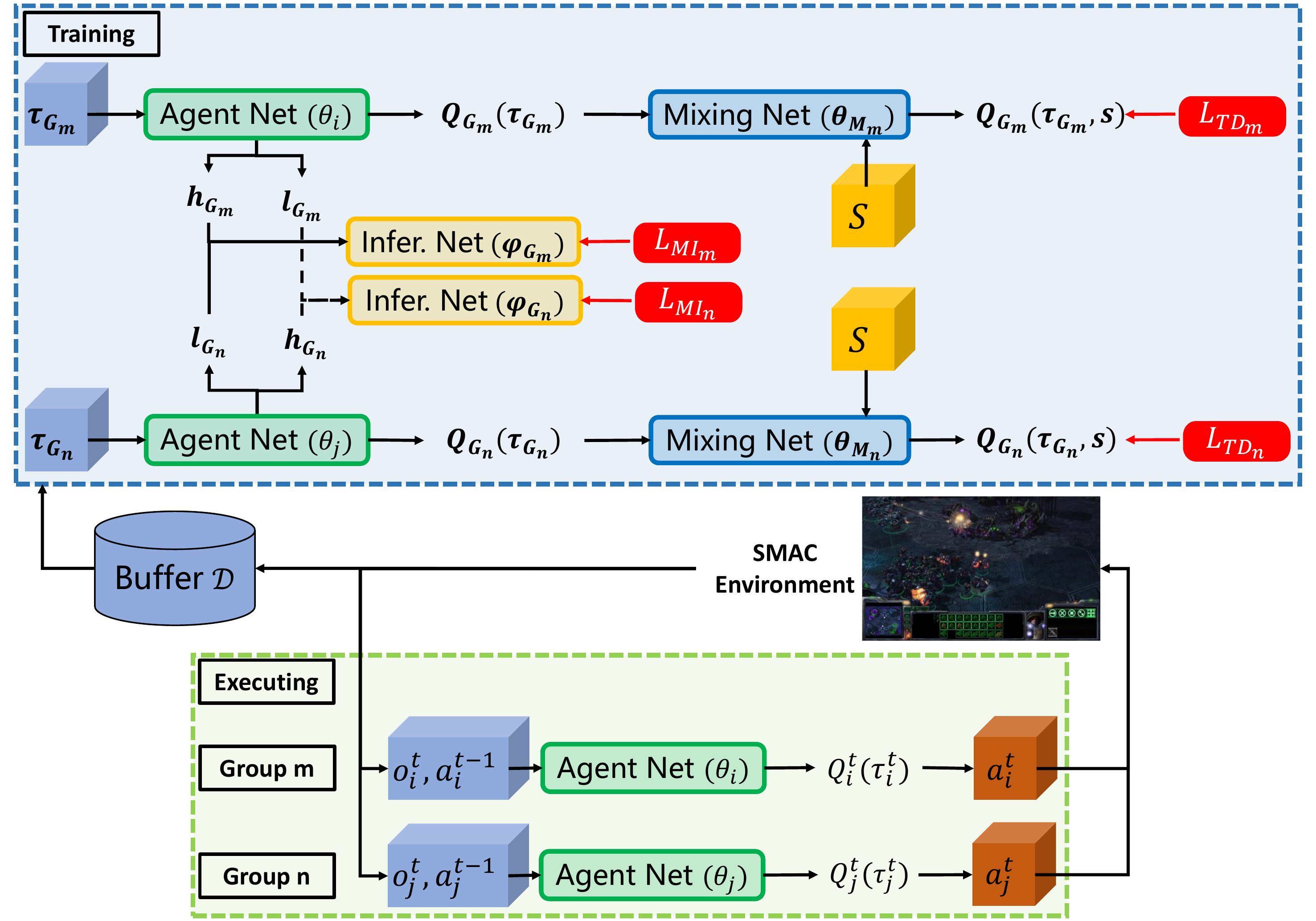}
\caption{An overview of data-stream of GHQ. During \textit{decentralized executing}, agent networks $\theta_i$ and $\theta_j$ generate $Q_i^t$ and $Q_j^t$ for choosing actions $a_i^t$ and $a_j^t$, respectively. The input of $\theta_i$ is the local observation $o_i^t$ and last action $a_i^{t-1}$ of agent $i$ in group $\mathcal{G}_m$. All necessary transition tuples $(\boldsymbol{\tau}, s, r)$ are stored into the replay buffer $\mathcal{D}$. During \textit{centralized training}, a batch of trajectories $\boldsymbol{\tau}_{\mathcal{G}_m}$ are sampled from $\mathcal{D}$ as the input of $\theta_i$ for calculating $\boldsymbol{Q}_{\mathcal{G}_m}(\boldsymbol{\tau}_{\mathcal{G}_m})$. Then, mixing network $\theta_{M_m}$ takes $\boldsymbol{Q}_{\mathcal{G}_m}(\boldsymbol{\tau}_{\mathcal{G}_m})$ and state $s$ for calculating $\boldsymbol{Q}_{\mathcal{G}_m}(\boldsymbol{\tau}_{\mathcal{G}_m}, \boldsymbol{s})$ and TD loss $\mathcal{L}_{TD_m}$. The GRU hidden states $h_{\mathcal{G}_m}$, $h_{\mathcal{G}_n}$ and the Gaussian distributions $l_{\mathcal{G}_m}$, $l_{\mathcal{G}_n}$ are generated from agent networks $\theta_i$ and $\theta_j$, and are used to calculate IGMI losses $\mathcal{L}_{MI_m}$ and $\mathcal{L}_{MI_n}$. Detailed formulas are shown in section \ref{IGMI-loss}.}
\label{fig-data_stream}
\end{figure*}

\subsection{Implementing details and Network Architecture}
Detailed network architecture is illustrated in Fig. \ref{fig-framework}, an overview of the data-stream of GHQ is illustrated in Fig. \ref{fig-data_stream}, and the pseudo-code of GHQ is given in Algorithm \ref{algo-GHQ}. As is shown in Fig. \ref{fig-framework} and \ref{fig-data_stream}, there are three kinds of networks marked with different colors. Agent network $\theta_{i}$ is marked in green and is shared by all agents $(i \in \mathcal{G}_m)$. $\theta_{i}$ receives the current observation $o_i^t$ and the last action $a_i^{t-1}$, and generates $Q_i^t$. The input is first sent to a \textit{Multi-Layer Perceptron} (MLP) and then a GRU layer. The hidden state of GRU $h_{\mathcal{G}_m}$ is sent to the following layers and the next time-step. The following layer is a Gaussian distribution layer generating $l_{\mathcal{G}_m}$ using $h_{\mathcal{G}_m}$, and then $l_{\mathcal{G}_m}$ is sampled and sent to the next two MLP layers. Eventually, a skip connection directly sends $h_{\mathcal{G}_m}$ to the final MLP layer, and $h_{\mathcal{G}_m}$ is concatenated with the output of formal MLP layer for generating $Q_i$.\par
Mixing network $\theta_{M_m}$ is marked in blue. It takes all $[Q_i]$ of the group $\mathcal{G}_m$ as input, and mixes with the state $s$ to produce the $Q_{\mathcal{G}_m}$. Four hyper-networks generate weights and bias $(w_1, b_1, w_2, b_2)$ with $s$, and only the absolute values of weights are used. The weights and bias multiply with the joint $[Q_i]$ procedurally and the intermediate results are activated to be non-negative, fulfilling the GIGM requirements. \par
Inference network $\psi_{\mathcal{G}_m}$ is marked in yellow and is only used to calculate $\mathcal{L}_{MI}$. It takes the hidden state of GRU $h_{\mathcal{G}_m}$ of group $\mathcal{G}_m$ and the Gaussian latent $l_{\mathcal{G}_n}$ of another group $\mathcal{G}_n$ as input. The input is first sent to an MLP layer and then a new Gaussian distribution layer to generate the inference distribution $q_{\mathcal{G}_m}(l_{\mathcal{G}_m} | l_{\mathcal{G}_n}, h_{\mathcal{G}_m})$. The MI-loss $\mathcal{L}_{MI}$ is calculated by the KL-divergence between the original distribution $p(l_{\mathcal{G}_m})$ and the inference distribution $q_{\mathcal{G}_m}(l_{\mathcal{G}_m} | l_{\mathcal{G}_n}, h_{\mathcal{G}_m})$.\par
Finally, when calculating the total loss $\mathcal{L}_{GHQ}$, adjusting weights $\lambda_{TD}$ and $\lambda_{MI}$ are introduced. In our implementation, we set $\lambda_{TD} = \lambda_{MI} =1$. We choose Adam \cite{kingma2014adam} as the optimizer, with the learning rate of all networks being 3e-4. The total training step is 5M and the maximum step for one episode is 200. The learning rate is scheduled to decay by multiplying the factor 0.5 every 50,000 episodes (averagely about 2M-3.5M steps). The reward discounting factor $\gamma$ is 0.99. The $\epsilon$ of the $\epsilon -greedy$ action selecting policy starts at 1.0, ends at 0.05 and linearly declines for 50,000 steps. The size of the memory buffer is 5,000 and the batch size is 32. A universal buffer saves all data for training, including trajectories of state $s^t$, observation $\boldsymbol{o}^t$, action $\boldsymbol{a}^t$ and reward $r^t$. After one episode, the latest data is inserted into the buffer and one batch of 32 episode data is sampled from the buffer and used for training. The following Table \ref{table-hyper-parameters} summarizes the hyper-parameters mentioned above. In addition, we use the latest version 4.10 of StarcraftII game on Linux to perform experiments, instead of the old version 4.6. \par

\begin{table}
\centering
\caption{The hyper-parameters of GHQ.}
\label{table-hyper-parameters}
\begin{tabular}{*{4}{c}}
\toprule
Hyper\_  &  \multirow{2}*{Value} & Hyper\_  &  \multirow{2}*{Value} \\
parameters &  & parameters &   \\
\midrule
\multirow{2}*{learning rate} & \multirow{2}*{3e-4} & learning rate & \multirow{2}*{0.5}\\
  &  &  annealing factor & \\
\multirow{2}*{training step} & \multirow{2}*{5M} & maximum step & \multirow{2}*{200}\\
  &  &  per episode & \\
$\gamma$ & 0.99 & $\epsilon$ & 1.0 $\to$ 0.05\\
memory buffer & 5000 & batch size & 32\\
\bottomrule
\end{tabular}
\end{table}

In summary, the total loss of GHQ is written as:

\begin{equation}
    \mathcal{L}_{GHQ}(\boldsymbol{\theta},\boldsymbol{\psi}) = \lambda_{TD} \mathcal{L}_{TD}(\boldsymbol{\theta}) + \lambda_{MI} \mathcal{L}_{MI}(\boldsymbol{\theta}, \boldsymbol{\psi})\ .
    \label{equa-Loss-GHQ}
\end{equation}

\section{Experiments and Results}\label{Experiments and Results}
\subsection{Designing New Asymmetric Heterogeneous Maps in SMAC}
In section \ref{LTH in SMAC}, we prove the existence of LTH in SMAC. However, the default setup of SMAC environment and default implementation of previous algorithms ignore the existence of LTH problem and the importance of asymmetric heterogeneous scenarios. \par
First, SMAC environment uses a padding vector to deal with the different interactive action-dim $|A_{act}|$. It increases the $|A_{act}|$ of $U_{spt}$ up to the $|A_{act}|$ of $U_{atk}$ with the padding vector, and masks unavailable actions when choosing. This solution covers up the existence of LTH problem. In addition, because of the padding vector, previous algorithms can apply parameter-sharing among all unit types. This implementation further prevents the MAS from learning better coordinating policy. In GHQ, all agents use their true $|A_{act}|$, and parameter-sharing is restricted between agents within the same group. \par
Second, it is ignored that the internal AI script of StarcraftII is incapable of coordinating and collaborating among multiple types of agents. As a result, the performance of enemies in symmetric heterogeneous maps is limited, and we consider that asymmetric heterogeneous maps are more fitted to perform and study the LTH problem. There are only two asymmetric heterogeneous maps in original SMAC maps: $3s5z\_vs\_3s6z$ and MMM2 (see Table \ref{table-SMAC-maps}). However, these two maps have shortages respectively. \par
All units in $3s5z\_vs\_3s6z$ are $U_{atk}$, while the difference is their \textit{shot-range} and \textit{health-point}. However, the heterogeneity of this map is restricted, because all $U_{atk}$ have the same ideal object. Algorithms can acquire high performance without any information about the types or other properties of agents. Another map, MMM2, contains Marine, Marauder, and Medivac (see Table \ref{table-SMAC-maps} and \ref{table-SMAC-units}). $U_{atk}$ and $U_{spt}$, ground unit and flying unit are all included in the map. However, since both sides contain all of the three types of units, the internal AI script is unable to perform well. Therefore, we need to design new asymmetric heterogeneous maps for experiments. \par
\begin{figure}
\centering
\includegraphics[width=0.99\columnwidth]{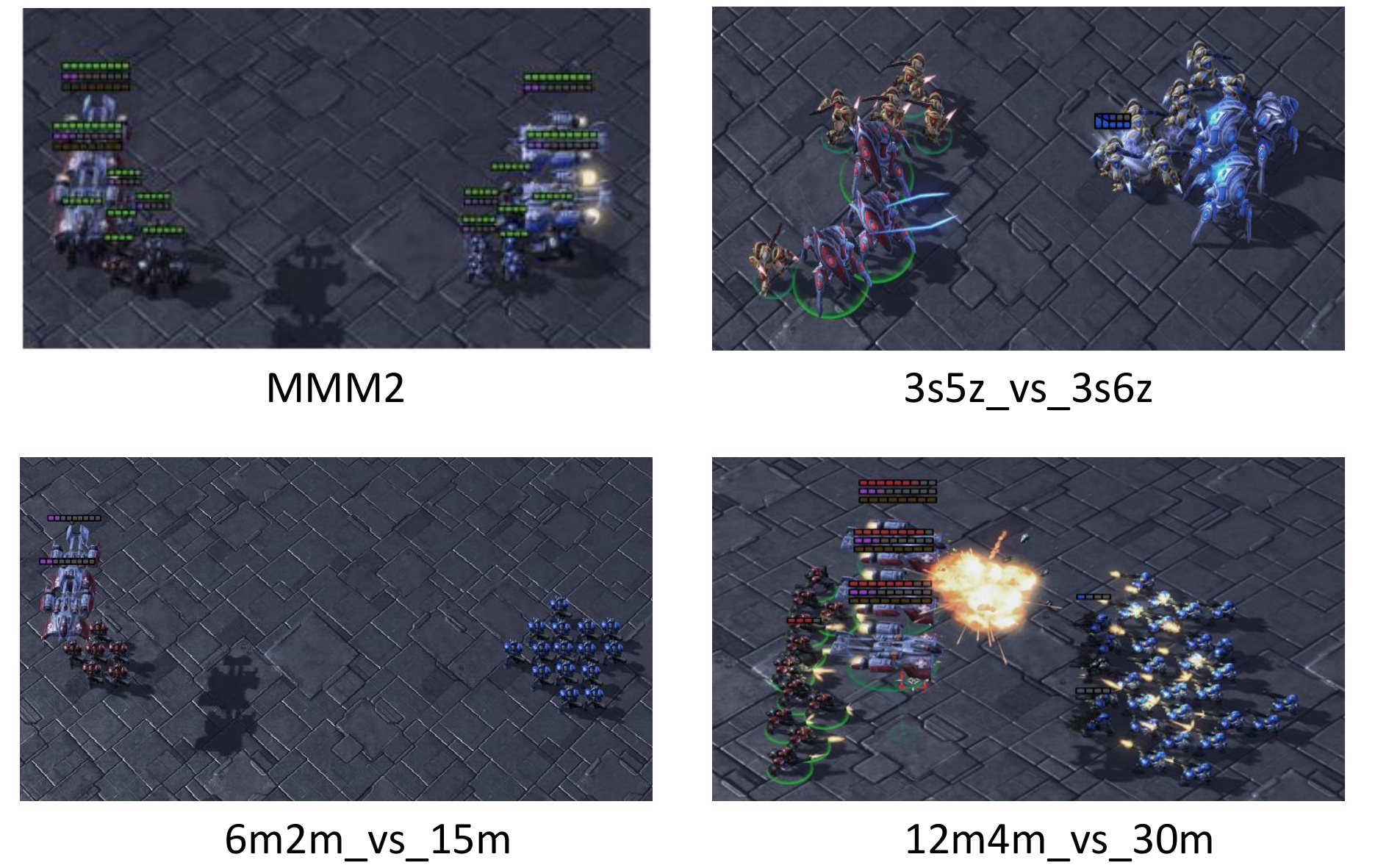}
\caption{Examples of SMAC maps. The lower two are ours.}
\label{fig-new-maps}
\end{figure}

Our maps, by contrast, avoid the shortages of original maps. For allies, we have Marine and Medivac, a $U_{atk}$ on the ground and a $U_{spt}$ in the air, which is similar to the common heterogeneous UAV-UGV MAS in \cite{robotic+ivic2020motion}. For enemies controlled by the internal AI script, we have only Marine to prevent the incapability of the script. We increase the number of enemy Marines to balance the difficulty of maps. Lots of pre-experiments are conducted to determine the specific number of all units. Table \ref{table-SMAC-new-maps} shows the information of all new maps. Fig. \ref{fig-new-maps} shows some examples of original and new maps.

\begin{table}
\centering
\caption{New Asymmetric Heterogeneous SMAC Maps.}
\label{table-SMAC-new-maps}
\setlength{\tabcolsep}{1.8mm}{
\begin{tabular}{*{5}{c}}
\toprule
\multirow{2}*{Map Name} &  Ally\_  &  Ally\_  &   Enemy\_  & \multirow{2}*{Difficulty} \\
 &  Marines  &  Medivacs  &  Marines  & \\
\midrule
6m2m\_15m & 6 & 2 & 15 & Easy\\
6m2m\_16m & 6 & 2 & 16 & Medium\\
8m3m\_21m & 8 & 3 & 21 & Medium\\
8m4m\_23m & 8 & 4 & 23 & Hard\\
12m4m\_30m & 12 & 4 & 30 & Ex-Hard\\
15m2m\_28m & 15 & 2 & 28 & Hard\\
16m2m\_30m & 16 & 2 & 30 & Ex-Hard\\
\bottomrule
\end{tabular}
}
\end{table}

\subsection{Environmental and Experimental Details}
In SMAC, all information provided by the environment is organized into tensors of pure data, all of which are either normalized into $[0,1]$ or transferred into one-hot vectors. We describe the necessary information details below for a better understanding of the SMAC environment. More details can be accessed in the official repository and source codes. \par
The state $S$ is only accessible by the mixing network $\theta_{M}$ during training. It consists of two major parts, \textit{ally-state} and \textit{enemy-state}:
\begin{itemize}
    \item \textit{ally-state} includes the percentage of \textit{health-point}, weapon cool-down timer, ally unit type, and absolute position of all allies;
    \item \textit{enemy-state} includes the percentage of \textit{health-point}, enemy unit type, and absolute position of all enemies.
\end{itemize}
\par
The observations $O$ is the input to the agent network $\theta_i$ for calculating $Q_i$. For agent $i$, the observation $o_i$ consists of four parts, \textit{moving-feature}, \textit{ally-feature}, \textit{enemy-feature} and \textit{own-feature}:
\begin{itemize}
    \item \textit{moving-feature} includes the ID of available moving action of agent $i$;
    \item \textit{ally-feature} includes the percentage of \textit{health-point}, unit type, relative distance, and relative position of other allies to agent $i$ within its \textit{sight-range}. Information about the agents out of the \textit{sight-range} of agent $i$ is not accessible;
    \item \textit{enemy-feature} includes the percentage of \textit{health-point}, unit type, relative distance, and relative position of all enemies to agent $i$ within its \textit{sight-range}. Information about the enemies out of the \textit{sight-range} of agent $i$ is not accessible;
    \item \textit{own-feature} includes the percentage of \textit{health-point} and unit type of agent $i$.
\end{itemize}
\par
As we have described in section \ref{LTF}, agent action consists of two parts: common-action $A_{com}$ and interactive-action $A_{act}$. The common action-dim $|A_{com}|$ is 6 for all agents. Action ID 0 is \textit{null} action only available for dead agents. Action ID 1 is \textit{stop} action, and ID 2, 3, 4, and 5 are \textit{moving} actions available for all living agents. The four \textit{moving} actions are pre-defined by the SMAC source codes, indicating moving up, down, left, and right with a certain \textit{moving\_amount} step-length. The interactive action-dim $|A_{act}|$ equals the number of interacting objects of a certain agent type. For $U_{atk}$, $|A_{act}|$ is the number of \textit{enemies}. For $U_{spt}$, $|A_{act}|$ is the number of \textit{allies}. \par
We use the default global dense reward function of SMAC. The MAS is rewarded when dealing \textit{damage} to the enemies, \textit{killing} enemies, and \textit{winning} the game. The \textit{damage} reward equals the value of the \textit{health-point} changes of enemies after one time-step, which is the absolute damage value dealt to the enemies. The \textit{killing} reward is 10 for every enemy-kill, and the \textit{winning} reward is 200 given at the terminal time-step. \par
We use the official implementations of all algorithms with minimal necessary adaptation to our new environmental settings. In general, we use the traditional winning-rate \textit{(WR)} as the measuring criterion. \textit{WR} is the probability of MARL agents eliminating all enemies and winning the game, and is approximated by the frequency of winning. We use the averaged \textit{WR} of 32 testing episodes. Testing episodes are taken every 10,000 training steps (about 1,000 training episodes). 5 rounds of complete experiments with different random seeds are performed for plotting the curve of the averaged \textit{WR} with the p-value being 0.05. As is shown in Fig. \ref{fig-framework} and \ref{fig-data_stream}, GHQ uses extra Inference networks to calculate IGMI loss. As a result, the computing time of GHQ is roughly about 1.5 times of the computing time of QMIX. Other value-based methods also consume more time than QMIX, indicating their more complexity than QMIX. \par

\begin{table}
\centering
\caption{The Enemy Strength \textit{(ES)}, Proportion of Supporting Units \textit{(POS)} and Winning-Rate \textit{(WR)} of QMIX-FT and GHQ on Homogeneous and Heterogeneous maps. The map name $XmYm\_Zm$ means that allies consist of X Marines and Y Medivacs while enemies consist of Z Marines.
}
\label{table-ES}
\setlength{\tabcolsep}{1.6mm}{
\begin{tabular}{*{5}{c}}
\toprule
Map Name & \textit{ES} & \textit{POS} & \textit{WR}(QMIX-FT) & \textit{WR}(GHQ)\\
\midrule
11m\_15m & 1.36 & / & 0.0 & / \\
12m\_15m & 1.25 & / & 0.5 & / \\
13m\_15m & 1.15 & / & 1.0 & / \\
15m\_20m & 1.33 & / & 0.0 & / \\
16m\_20m & 1.25 & / & 0.5 & / \\
17m\_20m & 1.18 & / & 1.0 & / \\
24m\_30m & 1.25 & / & 0.5 & / \\
25m\_30m & 1.20 & / & 0.9 & / \\
26m\_30m & 1.15 & / & 1.0 & / \\
6m2m\_15m & 2.50 & $25.0\%$ & 0.8 & 0.9 \\
6m2m\_16m & 2.67 & $25.0\%$ & 0.4 & 0.6 \\
6m2m\_17m & 2.83 & $25.0\%$ & 0.0 & 0.0 \\
7m2m\_15m & 2.14 & $22.2\%$ & 0.9  & 1.0\\
8m3m\_19m & 2.38 & $27.3\%$ & 0.9  & 1.0\\
8m3m\_21m & 2.63 & $27.3\%$ & 0.8 & 0.9 \\
15m2m\_28m & 1.87 & $11.8\%$ & 0.8 & 0.9 \\
16m2m\_28m & 1.75 & $11.1\%$ & 0.9  & 1.0\\
17m2m\_30m & 1.76 & $10.5\%$ & 0.9 & 0.9 \\
\bottomrule
\end{tabular}
}
\end{table}

\subsection{Criteria for Measuring Map Heterogeneity and Difficulty}
According to our analysis in section \ref{LTH in SMAC}, the existence of LTH in SMAC is clear. However, analyzing and quantifying the influence of LTH on agent policy is still required. Here, we propose objective criteria to measure the heterogeneity and difficulty of maps. \par
The \textit{Proportion of Supporting Units (POS)} is the proportion of the number of ally supporting units $|U_{spt_i}|$ divided by the number of overall ally units $|U_{i}|$. The \textit{Enemy Strength (ES)} is the ratio of weighted attacking units $U_{atk}$ of two sides, for measuring the strength of different $U_{atk}$. The result is calculated with the enemies divided by the allies.
\begin{equation}
\begin{aligned}
    POS &= \frac{|U_{spt_i}|}{|U_{i}|} = \frac{Ally\ Medivacs}{All\ Ally\ Units}  \ , \\
    ES &= \frac{\sum_{all\ types} w_e \cdot |U_{A_e}|}{\sum_{all\ types} w_i \cdot |U_{A_i}|} = \frac{Enemy\ Marines}{Ally\ Marines}  \ ,
\end{aligned}
\end{equation}
where $|U_{A_i}|$ and $|U_{A_e}|$ are the number of different types of attacking units $U_{atk}$ for allies and enemies, and $w_i$ and $w_e$ are the correction weights. \par
In our maps, since the only $U_{spt}$ is Medivac and the only $U_{atk}$ is Marine, \textit{POS} equals the proportion of Medivacs among all ally units. \textit{ES} equals the ratio of the number of Marines from two sides. It is obvious that high \textit{POS} represents high heterogeneity, because the high proportion of ally $U_{spt}$ indicates the serious influence introduced by the policy of $U_{spt}$. High \textit{ES} represents high difficulty, because the only way to win in SMAC is to control ally $U_{atk}$ eliminating all enemies, and high \textit{ES} indicates more enemy $U_{atk}$ than ally $U_{atk}$. \par
We design several homogeneous maps consisting of only Marine for both sides. The enemy consists of 15, 20, and 30 Marines, which is almost the same as our heterogeneous maps. The ally consists of Marines slightly less than the enemy (see Table \ref{table-ES}). According to the converged \textit{(WR)}, we conclude that in homogeneous maps with only Marines controlled by QMIX-FT \cite{riit+hu2021rethinking} algorithm, \textit{ES} and \textit{WR} are highly related and proportional. When \textit{ES} is about $1.25$, \textit{WR} is about 0.5; and when \textit{ES} is less than $1.18$, \textit{WR} keeps being $1.0$. Even if the total number of units is doubled, this relation remains unchanged. In symmetric homogeneous maps, \textit{ES} is at its minimum $1.0$, and thus it can be concluded that the MARL policy is easier to win than in asymmetric maps. \par
We further design additional heterogeneous maps (see Table \ref{table-ES}). On the one hand, the \textit{ES} of heterogeneous maps can be easily increased up to $1.7$ to $2.4$ in heterogeneous maps, when \textit{WR} of QMIX-FT is about $0.9$. Introducing heterogeneity into SMAC maps can significantly increase the difficulty of maps, so it is necessary to study and better utilize heterogeneity. On the other hand, \textit{POS} and \textit{ES} are highly related. In order to achieve high \textit{WR} in harder maps with high \textit{ES}, we need to increase \textit{POS} simultaneously with increasing attacking units. For example, in $6m2m\_16m$, \textit{ES} is 2.67 and \textit{POS} is $25.0\%$. Both GHQ and QMIX-FT can only achieve the \textit{WR} about $0.5$. By contrast, in $8m3m\_21m$, \textit{ES} is 2.63 and \textit{POS} is $27.3\%$, and the \textit{WR} reaches about $0.9$. \par
In conclusion, our results prove the shortage of original SMAC symmetric maps, and the ability of GHQ and QMIX-FT to handle the LTH problem with higher \textit{POS} and \textit{ES}. The following experiments show that better utilizing LTH helps GHQ to acquire higher \textit{WR} with smaller variance than QMIX-FT. Additionally, we conclude that the strength of 1 Medivac equals about $3.5$ Marines.

\subsection{Comparison Algorithms}
Experiments are taken in our seven new maps (see Table \ref{table-SMAC-new-maps}) and the MMM2 map as an original asymmetric heterogeneous map. We mainly choose value-based algorithms to run experiments for comparison, including vanilla QMIX \cite{qmix+rashid2018}, fine-tuned QMIX (QMIX-FT) \cite{riit+hu2021rethinking}, QPLEX \cite{wang2020qplex}, ROMA \cite{wang2020roma}, RODE \cite{wang2020rode}, MAIC \cite{yuan2022+maic} and CDS \cite{li2021+CDS}. We also run experiments for policy-based baseline algorithms, including COMA\cite{foerster2018+coma}, MAPPO\cite{mappo+yu2021surprising} and HAPPO\cite{happo+kuba2021trust}. \par
RODE and ROMA are \textit{role-based} algorithms, which learn and apply role policies online, end to end. These two algorithms are relatively similar to our group-based algorithms than others. However, ROMA can not learn effective policy within 5M (5 million) training-steps, because the default training step of ROMA is 20M. In RODE, several key hyper-parameters define the clustering and using of role-policies. The end-to-end clustering of role-policies makes it difficult to focus on the LTH property. Therefore, the performance of RODE is restricted. QPLEX, MAIC, and CDS modify the factorization structure of QMIX with distinct methods. COMA and MAPPO are actor-critic algorithms using the ``centralized critic decentralized actor'' (CCDA) architecture. These two algorithms apply parameter-sharing in actor networks and use one shared critic network. HAPPO uses independent network parameters for actor networks and proposes a monotonic policy-improving architecture with a theoretical guarantee.

\begin{figure*}
    \centering
	\includegraphics[width=0.99\textwidth]{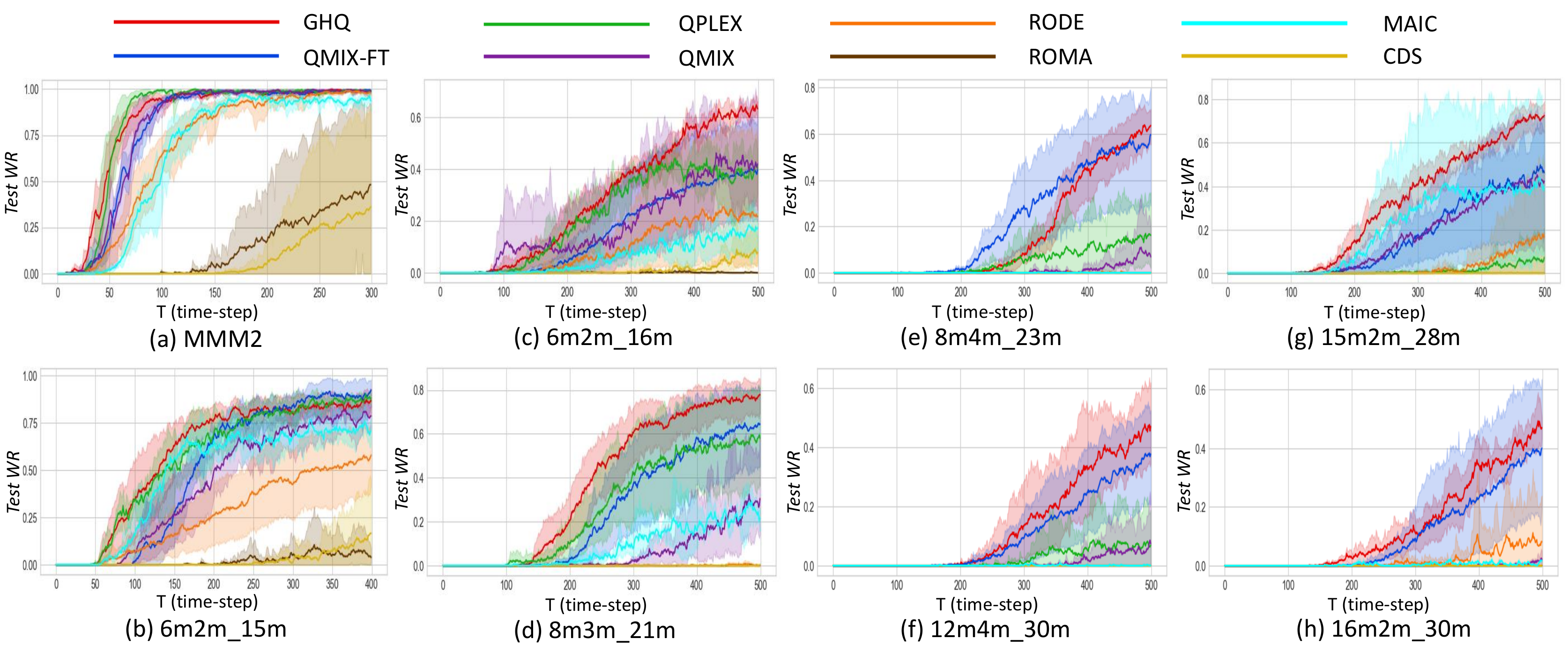}
	\caption{Results of Value-based Algorithms Comparison.}
	\label{fig-ALL8-baseline}
\end{figure*}

\subsection{Comparison Results}
The results of value-based algorithms are shown in section \ref{Value-based Algorithms Comparison}, Fig. \ref{fig-ALL8-baseline}, and Table \ref{table-value-based}. We criticize the performance of value-based algorithms with 4 groups of experiments. All of our GHQ results are in red color and the colors for other value-based algorithms are shown in the legend. The results of policy-based algorithms are shown in section \ref{Policy-based Algorithms Comparison} and Table \ref{table-policy-based}. Generally, all of the comparison algorithms suffer from the LTH problem and cannot acquire high \textit{WR} with small variance. Previous value-based algorithms are basically modified from QMIX and, to some extent, weaken the ability of QMIX to handle the LTH problem.

\subsubsection{Results of Value-based Algorithms Comparison}\label{Value-based Algorithms Comparison}

\begin{table*}
\centering
\caption{Results of Value-based Algorithms Comparison.}
\label{table-value-based}
\centering
\setlength{\tabcolsep}{1.7mm}{
\begin{tabular}{*{9}{c}}
\toprule
Map Name & GHQ & QMIX-FT & QMIX & QPLEX & RODE & ROMA & CDS & MAIC\\
\midrule
MMM2 & $\textbf{1.00}_{(0.00)}$ & $\textbf{1.00}_{(0.00)}$ & $\textbf{1.00}_{(0.00)}$ & $\textbf{1.00}_{(0.00)}$ & $\textbf{1.00}_{(0.00)}$ & $0.64_{(0.46)}$ & $0.42_{(0.40)}$ & $0.83_{(0.06)}$ \\
6m2m\_15m & $0.87_{(0.02)}$ & $\textbf{0.92}_{(0.02)}$ & $0.82_{(0.07)}$ & $0.89_{(0.04)}$ & $0.72_{(0.15)}$ & $0.06_{(0.09)}$ & $0.45_{(0.14)}$ & $0.67_{(0.06)}$\\
6m2m\_16m & $\textbf{0.62}_{(0.01)}$ & $0.42_{(0.10)}$ & $0.44_{(0.16)}$ & $0.39_{(0.08)}$ & $0.23_{(0.32)}$ & $0.00_{(0.00)}$ & $0.11_{(0.10)}$ & $0.20_{(0.09)}$ \\
8m3m\_21m & $\textbf{0.78}_{(0.08)}$ & $0.64_{(0.10)}$ & $0.40_{(0.19)}$ & $0.55_{(0.25)}$ & $0.01_{(0.02)}$ & $0.00_{(0.00)}$ & $0.00_{(0.00)}$ & $0.32_{(0.13)}$ \\
8m4m\_23m & $\textbf{0.62}_{(0.04)}$ & $0.60_{(0.11)}$ & $0.17_{(0.08)}$ & $0.21_{(0.13)}$ & $0.01_{(0.02)}$ & $0.00_{(0.00)}$ & $0.00_{(0.00)}$ & $0.00_{(0.00)}$ \\
12m4m\_30m & $\textbf{0.48}_{(0.09)}$ & $0.38_{(0.09)}$ & $0.02_{(0.02)}$ & $0.08_{(0.12)}$ & $0.00_{(0.00)}$ & $0.00_{(0.00)}$ & $0.00_{(0.00)}$ & $0.00_{(0.00)}$ \\
15m2m\_28m & $\textbf{0.72}_{(0.03)}$ & $0.48_{(0.17)}$ & $0.43_{(0.20)}$ & $0.08_{(0.12)}$ & $0.19_{(0.26)}$ & $0.00_{(0.00)}$ & $0.00_{(0.00)}$ & $0.42_{(0.31)}$ \\
16m2m\_30m & $\textbf{0.48}_{(0.04)}$ & $0.40_{(0.13)}$ & $0.00_{(0.00)}$ & $0.00_{(0.00)}$ & $0.06_{(0.09)}$ & $0.00_{(0.00)}$ & $0.00_{(0.00)}$ & $0.00_{(0.00)}$ \\
\bottomrule
\end{tabular}
}
\end{table*}

Results for value-based algorithms are shown in Fig. \ref{fig-ALL8-baseline} and Table \ref{table-value-based}. In Table \ref{table-value-based}, the results are the final \textit{WR}, and are averaged across 5 individual tests with different random-seeds. The standard deviations are followed. In Fig. \ref{fig-ALL8-baseline}, the lines and shadows are fitted across the whole data, so the values may be slightly different from the results in Table \ref{table-value-based}. \par
(1) \textit{We test all algorithms on the original asymmetric heterogeneous map MMM2.} The results are shown in Fig. \ref{fig-ALL8-baseline} (a). Because the map is relatively easy and almost all algorithms converge at 3M training steps, we only show the results ended at 3M steps for better presentation. The graph shows that \textit{WR} of most algorithms converged to 1.0 at about 1.5M steps with a relatively small variance. QPLEX and GHQ are slightly better than QMIX. RODE and MAIC converge at about 2.5M steps, which is slower than other algorithms. ROMA and CDS fail to converge at 3M steps. \par
(2) \textit{We decrease the heterogeneity of maps through decreasing \textit{POS}.} In Fig. \ref{fig-ALL8-baseline} (b), (d), (g), and (h), the number of Medivac remains to be 2, while the number of Marine is increased. Therefore, the \textit{POS} decreases from $25.0\%$ of (b), to $11.1\%$ of (h) (see Table \ref{table-ES}). As a result, algorithms using parameter-sharing among all agents learn better policy than the setting of increasing \textit{POS}. In general, algorithms perform well in small-scale maps (b) and (d), but only GHQ and QMIX-FT perform well in both of the large-scale maps (g) and (h). GHQ outperforms QMIX-FT with smaller variance. MAIC and QMIX perform well in (g) but fail in (h), indicating their limitation in handling large-scale problems. QPLEX and RODE cannot learn effective policy in (g) and (h), while ROMA and CDS completely fail in (g) and (h). RODE performs better in (d), (g), and (h) than in (c), (e), and (f), indicating the training of the role-selector requires homogeneous MARL settings. \par

\begin{figure*}
    \centering
    \includegraphics[width=0.99\textwidth]{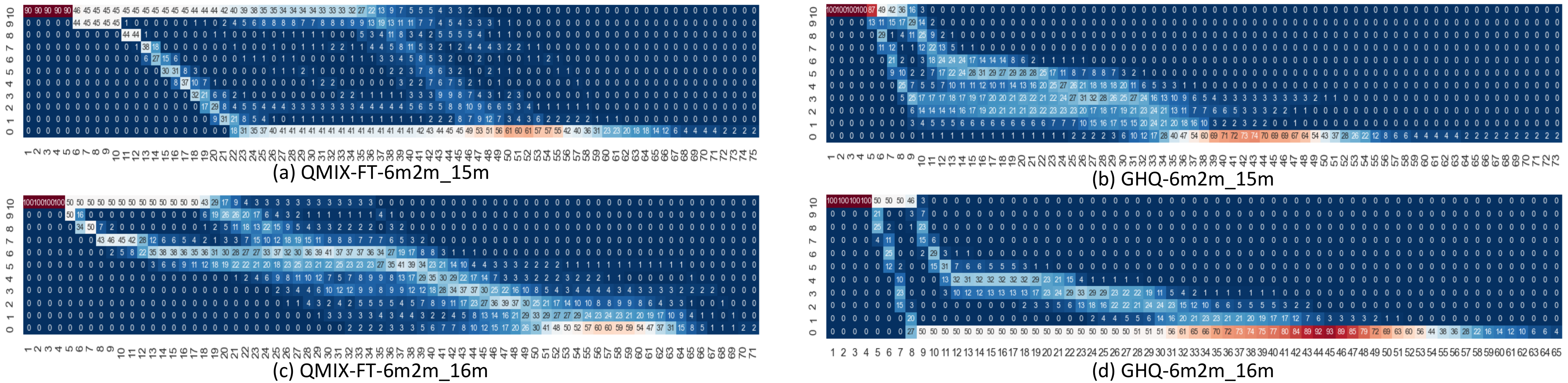}
    \caption{Heat-maps of $U_{spt}$' percentage of \textit{health-points} of GHQ and QMIX-FT in 6m2m\_15m and 6m2m\_16m.}
    \label{fig-extra-heat-map}
\end{figure*}

(3) \textit{We scale up all units of both sides simultaneously.} In Fig. \ref{fig-ALL8-baseline} (b) and (c), the \textit{POS} remains to be $25.0\%$, while the number increases from 2 to 4. Theoretically, the optimal policies of map (b) and (c) are similar. However, this scaling method combines the complexity of scalability and heterogeneity, making it harder for comparison algorithms to learn effective policies. In (b), most algorithms achieve high \textit{WR} within 5M steps, while GHQ converges fastest and RODE suffers from high variance and relatively low \textit{WR}. ROMA and CDS fail to learn effective policy in (b). However, in (c), almost all comparison algorithms fail to learn effective policy. GHQ and QMIX-FT outperform other algorithms and have not yet converged at 5M steps. QPLEX also suffers from complexity, but generally performs better than QMIX, ROMA, RODE, MAIC, and CDS. \par

\begin{sidewaystable}
\centering
\caption{Results of Independent t-test of GHQ against other Value-based Algorithms.}
\label{table-value-based-ttest}
\centering
\begin{tabular}{*{8}{c}}
\toprule
Map Name & QMIX-FT & QMIX & QPLEX & RODE & ROMA & CDS & MAIC\\
\midrule
MMM2 & $\textbf{0.268}_{(0.788)}$ & $\textbf{0.268}_{(0.788)}$ & $\textbf{0.268}_{(0.788)}$ & $\textbf{0.268}_{(0.788)}$ & $\textbf{36.584}_{(0.000)}$ & $\textbf{81.881}_{(0.000)}$ & $\textbf{993.291}_{(0.000)}$ \\
6m2m\_15m & $-1962.959_{(0.000)}$ & $\textbf{225.391}_{(0.000)}$ & $-278.309_{(0.000)}$ & $\textbf{143.981}_{(0.000)}$ & $\textbf{218.205}_{(0.000)}$ & $\textbf{479.887}_{(0.000)}$ & $\textbf{1202.379}_{(0.000)}$\\
6m2m\_16m & $\textbf{435.475}_{(0.000)}$ & $\textbf{155.590}_{(0.000)}$ & $\textbf{829.987}_{(0.000)}$ & $\textbf{82.276}_{(0.000)}$ & $\textbf{140625.813}_{(0.000)}$ & $\textbf{1141.160}_{(0.000)}$ & $\textbf{1128.510}_{(0.000)}$ \\
8m3m\_21m & $\textbf{259.927}_{(0.000)}$ & $\textbf{229.853}_{(0.000)}$ & $\textbf{85.563}_{(0.000)}$ & $\textbf{2723.414}_{(0.000)}$ & $\textbf{2764.715}_{(0.000)}$ & $\textbf{2764.716}_{(0.000)}$ & $\textbf{557.180}_{(0.000)}$ \\
8m4m\_23m & $\textbf{35.853}_{(0.000)}$ & $\textbf{1516.924}_{(0.000)}$ & $\textbf{558.173}_{(0.000)}$ & $\textbf{8364.497}_{(0.000)}$ & $\textbf{8789.821}_{(0.000)}$ & $\textbf{8789.822}_{(0.000)}$ & $\textbf{8789.822}_{(0.000)}$ \\
12m4m\_30m & $\textbf{194.167}_{(0.000)}$ & $\textbf{1286.745}_{(0.000)}$ & $\textbf{557.278}_{(0.000)}$ & $\textbf{1344.418}_{(0.000)}$ & $\textbf{1344.418}_{(0.000)}$ & $\textbf{1344.418}_{(0.000)}$ & $\textbf{1344.418}_{(0.000)}$ \\
15m2m\_28m & $\textbf{180.829}_{(0.000)}$ & $\textbf{160.417}_{(0.000)}$ & $\textbf{1024.268}_{(0.000)}$ & $\textbf{169.331}_{(0.000)}$ & $\textbf{18146.448}_{(0.000)}$ & $\textbf{18146.450}_{(0.000)}$ & $\textbf{68.470}_{(0.000)}$ \\
16m2m\_30m & $\textbf{102.772}_{(0.000)}$ & $\textbf{6805.080}_{(0.000)}$ & $\textbf{6805.081}_{(0.000)}$ & $\textbf{1100.550}_{(0.000)}$ & $\textbf{6805.080}_{(0.000)}$ & $\textbf{6805.081}_{(0.000)}$ & $\textbf{6805.081}_{(0.000)}$ \\
\bottomrule
\end{tabular}
\end{sidewaystable}

(4) \textit{We increase the heterogeneity of maps through increasing \textit{POS}.} In Fig. \ref{fig-ALL8-baseline} (d), (e), and (f), the \textit{POS} are $25.0\%$, $27.2\%$ and $33.3\%$, while the number are 2, 3, and 4. The results show that GHQ achieves the best results in all maps with relatively small variance, indicating the effectiveness of the IOG method and IGMI loss. IOG allows different types of agents to possess different network parameters and thus reduces the influence of increasing Medivacs. IGMI loss helps to increase the correlation between group trajectories and thus increases cooperation between groups. QMIX-FT achieves relatively similar \textit{WR} to GHQ against other algorithms, but still suffers from high variance. QPLEX performs well in (d) and (e), but the \textit{WR} decreases evidently in (f), indicating the influence of LTH. \textit{WR} of RODE is about 0.2 in (d), but remains zero in other maps. ROMA fails to learn effective policy in all maps. MAIC achieves a \textit{WR} of about 0.2 in (d) and (e), but fails in (f). CDS has little \textit{WR} in (d) but fails in other maps. \par

\subsubsection{Independent t-test and further analysis of GHQ against other Value-based Algorithms}
In SMAC, we cannot conduct the experiment of two algorithms attack against each other. Therefore, we cannot directly count the win-lose relationship between algorithms for the statistical tests in \cite{stat-derrac2011practical}. As an alternative, we use the data in Table \ref{table-value-based} to conduct independent t-tests between GHQ and other value-based algorithms to prove the significance of the obtained results. We assume that the distributions of all results are normal, following the mean values and the standard deviations in the table. We use Scipy to generate the distributions with the size being 500, and then run the independent t-tests. The results of t-tests are shown in Table \ref{table-value-based-ttest}. The t-statistics are shown in the table with the p-values followed. \par
It is obvious that almost all p-values are smaller than 0.05, indicating the significance of the results. Only the p-values of the results of GHQ against QMIX-FT, QMIX, QPLEX, and RODE in MMM2 are greater than 0.05, indicating that the result of GHQ has no significant difference against the result of the 4 algorithms, which is proved by Table \ref{table-value-based}. The t-statistics are also almost all positive, indicating the superior performance of GHQ against other algorithms. \par
In 6m2m\_15m, the mean value of GHQ is smaller than QMIX-FT and QPLEX. First, we need to point out that, as is shown in Fig. \ref{fig-ALL8-baseline}(b), the curve of the \textit{WR} of GHQ grows faster than the other two algorithms, indicating the faster learning speed of GHQ. Second, for further analysis, we draw a heat-map of $U_{spt}$' percentage of \textit{health-points} following the method in section \ref{Visualization Analysis of Trained Policies}, and the result is shown in Fig. \ref{fig-extra-heat-map}. It can be concluded that GHQ learns similar policies in 6m2m\_15m and 6m2m\_16m, which is to ``let $U_{spt}$ take damage for preserving $U_{atk}$''. However, even though QMIX-FT manages to learn a similar policy with GHQ in 6m2m\_15m, it fails to learn the proper policy in 6m2m\_16m. This phenomenon indicates the increasing difficulty of 6m2m\_16m than 6m2m\_15m, as the optimal policy becomes harder to learn. \par

\subsubsection{Results of Policy-based Algorithms Comparison}\label{Policy-based Algorithms Comparison}
Due to the discrete property of SMAC, value-based algorithms generally have achieved better results than policy-based algorithms \cite{qmix+rashid2018,riit+hu2021rethinking,mappo+yu2021surprising}. To support this conclusion, we conduct experiments of COMA, MAPPO, and HAPPO against GHQ and QMIX-FT. Results for these algorithms are shown in Table \ref{table-policy-based}. The final \textit{WRs} are shown in the table, and the results are averaged across 3 individual tests with different random-seeds and the standard deviations are followed. Original papers of MAPPO and HAPPO run experiments in SMAC for 10M training-steps, so we list the results of 5M and 10M training-steps separately. COMA can only acquire \textit{WR} in MMM2 and fail in all other maps. MAPPO performs the best among the 3 policy-based algorithms, especially in 6m2m\_15m, 8m4m\_23m, and 12m4m\_30m. These maps have relatively high \textit{ES} and \textit{POS}, indicating the potential of MAPPO handling LTH problems. HAPPO performs worse than other policy-based algorithms. One possible reason is that HAPPO implements Multi-Agent Advantage Decomposition (MAAD) via the random sequential update and execute scheme. However, in the LTH problem, the sequential partial order of agent actions can significantly affect the final joint policy. In conclusion, the results show that value-based algorithms generally perform better than policy-based algorithms, and GHQ outperforms all policy-based baseline algorithms.

\begin{table*}
\centering
\caption{Results of Policy-based Algorithms Comparison.}
\label{table-policy-based}
\centering
\begin{tabular}{*{8}{c}}
\toprule
Map Name & COMA & MAPPO & MAPPO & HAPPO & HAPPO & QMIX-FT & GHQ\\
Training steps & 5M & 5M & 10M & 5M & 10M & 5M & 5M \\
\midrule
MMM2 & $0.63_{(0.14)}$ & $0.72_{(0.23)}$ & $0.84_{(0.09)}$ & $0.05_{(0.02)}$ & $0.10_{(0.02)}$ & $\textbf{1.00}_{(0.00)}$ & $\textbf{1.00}_{(0.00)}$ \\
6m2m\_15m & $0.00_{(0.00)}$ & $0.18_{(0.31)}$ & $0.15_{(0.25)}$ & $0.00_{(0.00)}$ & $0.00_{(0.00)}$ & $\textbf{0.92}_{(0.02)}$ & $0.87_{(0.02)}$ \\
6m2m\_16m & $0.00_{(0.00)}$ & $0.00_{(0.00)}$ & $0.00_{(0.00)}$ & $0.00_{(0.00)}$ & $0.00_{(0.00)}$ & $0.42_{(0.10)}$ & $\textbf{0.62}_{(0.01)}$ \\
8m3m\_21m & $0.00_{(0.00)}$ & $0.06_{(0.11)}$ & $0.00_{(0.00)}$ & $0.04_{(0.01)}$ & $0.07_{(0.01)}$ & $0.64_{(0.10)}$ & $\textbf{0.78}_{(0.08)}$ \\
8m4m\_23m & $0.00_{(0.00)}$ & $0.22_{(0.20)}$ & $0.55_{(0.05)}$ & $0.00_{(0.00)}$ & $0.00_{(0.00)}$ & $0.60_{(0.11)}$ & $\textbf{0.62}_{(0.04)}$ \\
12m4m\_30m & $0.00_{(0.00)}$ & $0.10_{(0.08)}$ & $0.21_{(0.13)}$ & $0.00_{(0.00)}$ & $0.00_{(0.00)}$ & $0.38_{(0.09)}$ & $\textbf{0.48}_{(0.09)}$ \\
15m2m\_28m & $0.00_{(0.00)}$ & $0.00_{(0.00)}$ & $0.00_{(0.00)}$ & $0.00_{(0.00)}$ & $0.00_{(0.00)}$ & $0.48_{(0.17)}$ & $\textbf{0.72}_{(0.03)}$ \\
16m2m\_30m & $0.00_{(0.00)}$ & $0.00_{(0.00)}$ & $0.00_{(0.00)}$ & $0.00_{(0.00)}$ & $0.00_{(0.00)}$ & $0.40_{(0.13)}$ & $\textbf{0.48}_{(0.04)}$ \\
\bottomrule
\end{tabular}
\end{table*}

\subsection{Ablation Study}

\begin{figure}
    \centering
    \includegraphics[width=0.99\columnwidth]{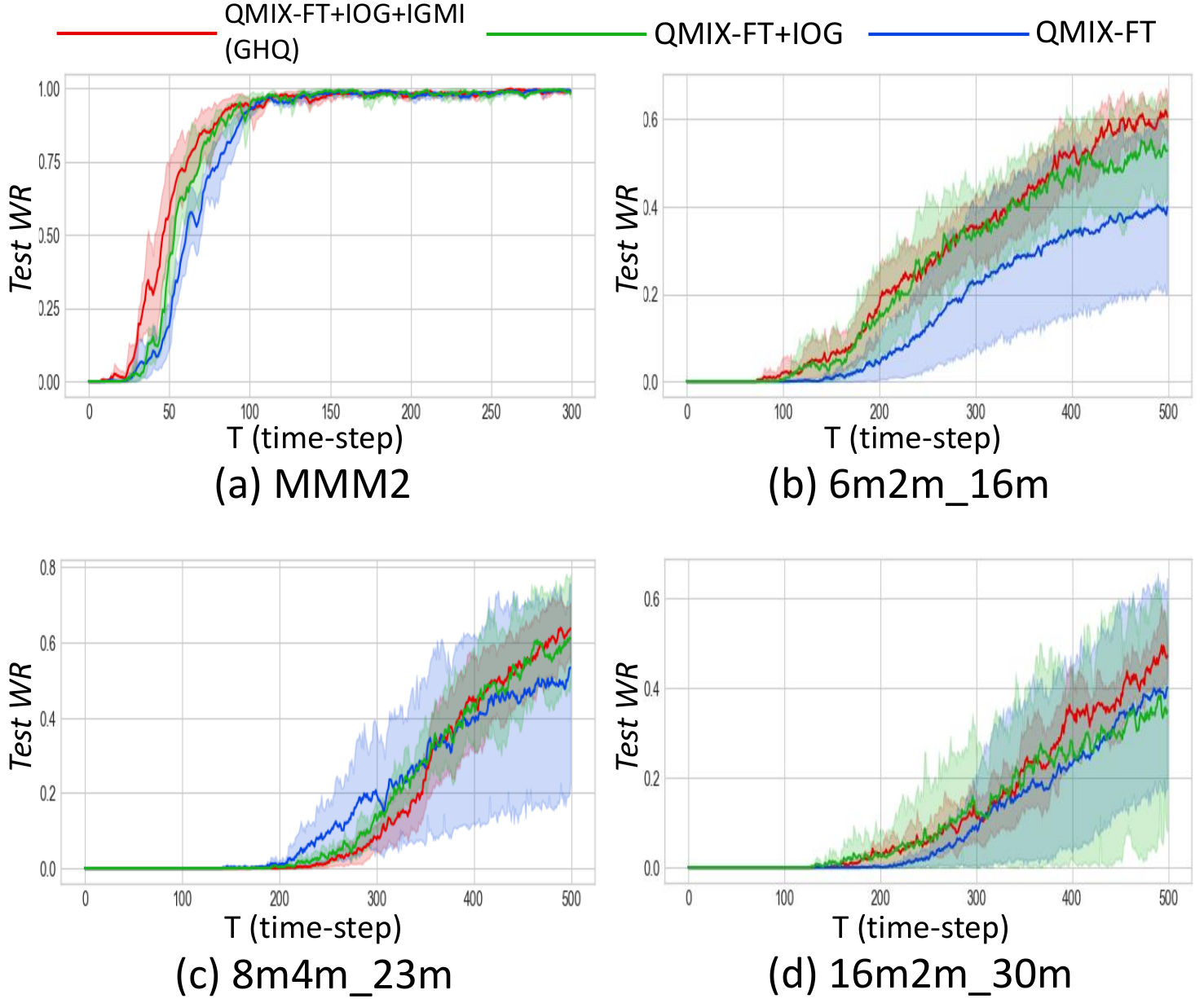}
    \caption{Results for ablation tests about IOG and IGMI.}
    \label{fig-abaltion}
\end{figure}

The ablation study consists of two experiments. \ref{Ablation Tests} is the ablation test about two component parts of GHQ, IOG, and IGMI. Because the IGMI must be applied between two agent groups, it is incapable of testing ``QMIX-FT+IGMI'' individually. Therefore, 3 groups of ablation tests are taken in 4 maps, as is shown in Fig. \ref{fig-abaltion}. Another experiment in \ref{Visualization Analysis of Trained Policies} is the visualization analysis of trained policies about GHQ and QMIX-FT in 6m2m\_16m. We visualize the trained policies of the two algorithms in heat-maps to show the influence of IOG and IGMI on policy learning. The temperature of heat-maps is the counting sum of corresponding agents. The results are shown in Fig. \ref{fig-heat-map}.

\subsubsection{Ablation Tests about IOG and IGMI}\label{Ablation Tests}
In order to analyze the effectiveness of IOG method and IGMI loss in different maps, we take ablation tests in (a) MMM2, (b) 6m2m\_16m, (c) 8m4m\_23m, and (d) 16m2m\_30m. QMIX-FT and QMIX-FT+IOG are the ablation groups. The results are shown in Fig. \ref{fig-abaltion}. \par
In general, as our expectation, IOG method helps to improve the performance of QMIX-FT, and IGMI loss helps to reduce variance. Fig. \ref{fig-abaltion} (a) shows that all algorithms are able to conquer the MMM2 map within 1.5M steps, while QMIX-FT+IOG and GHQ are converged slightly faster than QMIX-FT. In (b) 6m2m\_16m, IOG and IGMI are performing well. They not only improve the \textit{WR}, but also reduce the variance. In (c) 8m4m\_23m, the \textit{WR} of QMIX-FT increases faster than the other two algorithms before 3M steps. But IOG method manages to find a good cooperating policy, and converges to a better \textit{WR} at 5M steps with smaller variance. A higher derivative of IOG method at 3M to 4M steps indicates the progress of learning better policy. In (d) 16m2m\_30m, GHQ outperforms QMIX-FT with higher \textit{WR} and smaller variance. QMIX-FT+IOG receives a similar result with QMIX-FT, but has even larger variance. The main reason is that the difference between two groups is so large. Introducing IGMI loss helps to restrict the difference and improve the correlation between groups. Therefore, GHQ achieves the best result among the three testing algorithms.

\subsubsection{Visualization Analysis of Trained Policies}\label{Visualization Analysis of Trained Policies}

\begin{figure*}
\centering
\includegraphics[width=0.99\textwidth]{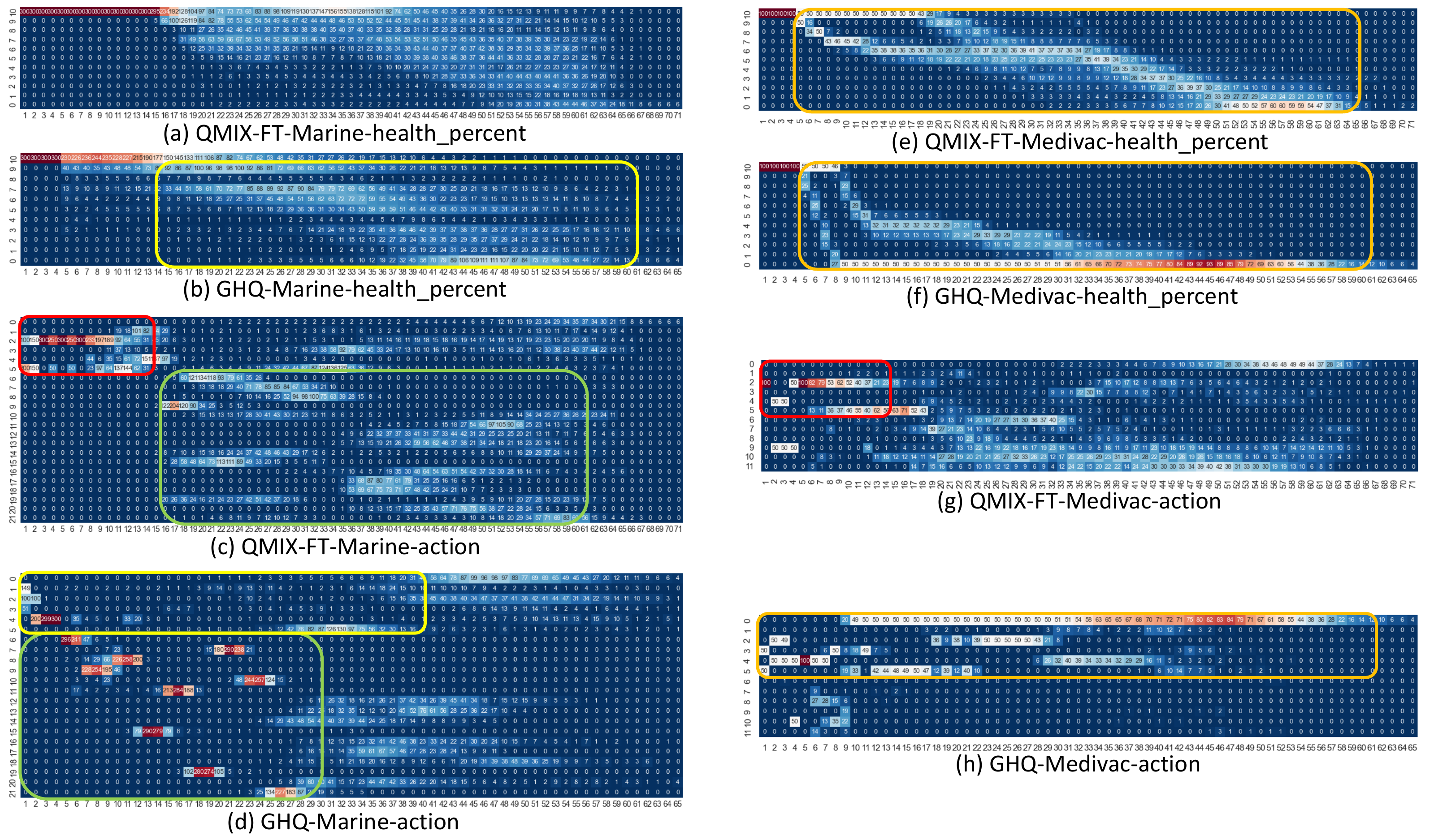}
\caption{Heat-maps for the trained policies of GHQ and QMIX-FT in 6m2m\_16m.}
\label{fig-heat-map}
\end{figure*}

We choose the trained policies of GHQ and QMIX-FT in 6m2m\_16m at 5M training steps for analysis. The two algorithms are implemented with the same hyper-parameter and similar network capacity. The differences in choosing action and \textit{health-points} of the two unit types controlled by GHQ and QMIX-FT are visualized in Fig. \ref{fig-heat-map}. We test the two policies 50 times and record their trajectories. We calculate the sum of agents' \textit{chosen-actions} and agents' percentage of \textit{health-points}, and visualize them in heat-maps of Fig. \ref{fig-heat-map}. The horizontal coordinates of all heat-maps are the time-step $T$, and the vertical coordinates are the every $10^{th}$ percentile of \textit{health-points} in (a), (b), (e) and (f), or the action ID number in (c), (d), (g) and (h). The temperature is the counting sum of corresponding agents at certain time-step with certain status. In action heat-maps, action ID 0 to 5 are common actions $A_{com}$ for moving and stopping, while the rest are interactive actions $A_{act}$ for attacking or healing. The chosen network parameters of GHQ and QMIX-FT reach the same \textit{WR} of about 0.8 after 5M training steps, noting that GHQ learns faster than QMIX-FT. Fig. \ref{fig-heat-map} clearly shows that the two algorithms achieve similar results through different agent policies. We observe three key phenomena. \par
(1) \textit{Parameter-sharing among different agent types do influence agent policy.} As is suggested in \cite{happo+kuba2021trust}, parameter-sharing restricts network parameters from being diverse. Red boxes in Fig. \ref{fig-heat-map} (c) and (g) show a similar policy pattern of ``first move and then stop to attack/heal'' for two types of agents in QMIX-FT. Specifically, both types of agents prefer to choose action 2 and 5 in the first 14 time-steps. In GHQ, however, the diversity of different groups is guaranteed, as is generally shown in (d) and (h). In addition, comparing the Medivac policy of QMIX-FT and GHQ in (g) and (h), it is clear that the QMIX-FT Medivac policy in (g) is more similar to the QMIX-FT Marine policy in (c) than the GHQ policies in (h) and (d). \par
(2) \textit{GHQ improves group policy learning.} Green boxes in Fig. \ref{fig-heat-map} (c) and (d) indicate that Marine controlled by GHQ learns better ``focusing and firing'' policy, as the temperature of $A_{act}$ are notably hotter than QMIX-FT. GHQ agents learn to focus and fire at one specific enemy target within several time-steps, which makes them quickly eliminate enemies and reduces their damage. By contrast, QMIX-FT agents learn to fire at several targets at the same time, which reduces the speed of elimination and causes more damage. Yellow box in Fig. \ref{fig-heat-map} (d) shows that the moving policies of GHQ Marines are also significantly different from QMIX-FT. GHQ Marines finish their movement in the first 4 time-steps with decisive actions and form a tight front. They tend to stay together and therefore take enemy damage simultaneously, which leads to a similar decreasing tendency of \textit{health-point} and the two obvious temperature valleys at 80 and 40 percentile in the yellow box of (b). \par
(3) \textit{GHQ improves inter-group cooperating.} Orange boxes in Fig. \ref{fig-heat-map} (e) and (f) represent the decreasing curves of the \textit{health-point} of Medivacs. GHQ Medivacs learns a better ``distracting'' policy than QMIX-FT Medivacs. One GHQ Medivac first moves toward enemies and attracts fire to prevent enemies from attacking ally Marines. This policy is proved by the orange box in (h) with the ``action 0 line'', indicating the death of one Medivac agent. Then, the other Medivac agent moves on to keep attracting enemy fire. As a result, the figure in (f) consists of two independent curves. The distraction policy performed by GHQ Medivacs is a fabulous tactic and significantly differs from the policies of GHQ Marines, indicating that GHQ is capable of utilizing LTH for better cooperation.

\section{Conclusion}\label{Conclusion}
In this paper, we focus on the cooperative heterogeneous MARL problem, especially the asymmetric heterogeneous MARL problems. In order to describe and study the heterogeneous MARL problem, we propose the \textit{Local Transition Heterogeneity} (LTH) with a formal definition. To support the definition of LTH, we first define the Local Transition Function (LTF) and several auxiliary concepts. Furthermore, we study the existence and influence of LTH in SMAC. \par
In order to primarily solve the LTH problem, we first propose the Grouped Individual-Global-Max (GIGM) consistency. Following the restriction of GIGM, we further propose the Ideal Object Grouping (IOG), the Inter-Group Mutual Information (IGMI) loss, and the \textit{hybrid} factorization structure. The combination of these three methods is our novel Grouped Hybrid Q-learning (GHQ) algorithm. Experiments are conducted in asymmetric heterogeneous SMAC maps to show that GHQ outperforms other state-of-the-art algorithms. The results prove the necessity to study and utilize LTH for studying more complex scenarios in SMAC. \par
We believe that the study of heterogeneity is indispensable for future MARL studies, and we hope that the mathematical definitions and analysis can help future studies on the cooperative heterogeneous MARL problem. Due to the restriction of computing resources and network structure, we are unable to study large-scale problems or transfer learning problems in heterogeneous MARL. In the future, we will try to solve more large-scale and complex heterogeneous MARL problems in other maps and environments.

\section*{Statements and Declarations}
\begin{itemize}
\item Funding: No funding was received to assist with the preparation of this manuscript.
\item Competing interests: The authors have no competing interests to declare that are relevant to the content of this article.
\item Ethics approval: This article does not involve any ethical problem which needs approval.
\item Consent to participate: All authors have seen and approved the final version of the manuscript being submitted.
\item Consent for publication: All authors warrant that the article is our original work, hasn't received prior publication, and isn't under consideration for publication elsewhere. A preprint version of our manuscript has been submitted to arXiv, and the page is \url{https://arxiv.org/abs/2303.01070}. The journal version improves the overall structure of the article, and enhances with more definitions, demonstrations, and experiments.
\item Availability of data and materials: The datasets generated during and/or analyzed during the current study are available from the corresponding author on reasonable request.
\item Code availability: The codes for this article are available from the corresponding author on reasonable request.
\item Authors' contributions: Conceptualization: [Xiaoyang Yu, Kai Lv, Xiangsen Wang]; Methodology: [Xiaoyang Yu, Kai Lv]; Formal analysis and investigation: [Xiaoyang Yu]; Writing - original draft preparation: [Xiaoyang Yu]; Writing - review and editing: [Xiaoyang Yu, Youfang Lin, Xiangsen Wang, Sheng Han, Kai Lv]; Funding acquisition: [Youfang Lin, Sheng Han]; Resources: [Youfang Lin, Sheng Han]; Supervision: [Youfang Lin, Sheng Han, Kai Lv].
\end{itemize}

\makeatletter
\renewcommand\@biblabel[1]{#1.}
\makeatother
\bibliography{GHQ}

\end{document}